\documentclass{article}

\usepackage{authblk}
\usepackage[font=small]{caption}
\usepackage{graphicx}
\usepackage{epstopdf}
\usepackage{amsmath}
\usepackage{amssymb}
\usepackage{dsfont}
\usepackage[hidelinks]{hyperref}
\newcommand{\C}{\mathbb{C}}
\newcommand{\im}[1]{\:\text{Im}\left({#1}\right)} 
\newcommand{\re}[1]{\:\text{Re}\left({#1}\right)} 
\newcommand{\E}[1]{\mathds{E}\left\{{#1}\right\}} 
\newcommand{\mat}[1]{\begin{bmatrix} #1\end{bmatrix}} 

\newcommand{\norm}[1]{\left\Vert{ #1 }\right\Vert}

\DeclareMathOperator*{\argmin}{arg\,min}
\let\citeleft=(
\let\citeright=)

\begin{document}


\pdfinfo{
   /Author (Siddharth Iyer, Frank Ong, Kawin Setsompop, Mariya Doneva and Michael Lustig)
   /Title (SURE-based Automatic Parameter Selection For ESPIRiT Calibration)
}

\title{\vspace{-2cm}SURE-based Automatic Parameter Selection For ESPIRiT Calibration}

\author[1,2,4]{Siddharth Iyer}
\author[1]{Frank Ong}
\author[2]{Kawin Setsompop}
\author[3]{Mariya Doneva}
\author[1]{Michael Lustig}

\affil[1]{\small Department of Electrical Engineering and Computer Science, University of California, 
          Berkeley, California, 94720, United States of America}
\affil[2]{\small Athinoula A. Martinos Center for Biomedical Imaging, Department of Radiology, Massachusetts General
          Hospital, Charlestown, Massachusetts, 02129, United States of America}
\affil[3]{\small Philips Research Laboratories, Hamburg, Germany}
\affil[4]{\small Department of Electrical Engineering and Computer Science, Massachusetts Institute of Technology, 
          Cambridge, Massachusetts, 02139, United States of America}
\date{}
\maketitle

\vfill
\noindent
\textit{Running head:} SURE-Calibrated ESPIRiT.

\noindent
\textit{Address correspondence to:} \\
  Michael Lustig\\
  506 Cory Hall\\
  University of California, Berkeley\\
  Berkeley, CA 94720\\
  United States of America \\
  mlustig@eecs.berkeley.edu

\noindent
This work was supported by NIH R01EB019241, NIH R01EB009690, NSF Graduate Fellowship, U24EB029240, R01HL136965, R01EB026136, NIH R01EB020613, R01 MH116173

\noindent
Approximate word count: 200 (Abstract) 5000 (body)\\

\noindent
Part of this work has been presented at the ISMRM Annual Conference 2016.

\noindent
Published at \textit{Magnetic Resonance in Medicine} as a Full Paper.

\clearpage

\section{Abstract}

\noindent
\textbf{Purpose}: ESPIRiT is a parallel imaging method that estimates coil sensitivity maps from the auto-calibration
region (ACS). This requires choosing several parameters for the optimal map estimation.
While fairly robust to these parameter choices, occasionally, poor selection can result in reduced performance.
The purpose of this work is to automatically select parameters in ESPIRiT for more robust and consistent performance
across a variety of exams.

\noindent
\textbf{Methods}:
By viewing ESPIRiT as a denoiser, Stein's unbiased risk estimate (SURE) is leveraged to automatically optimize
parameter selection in a data-driven manner.
The optimum parameters corresponding to the minimum true squared error, minimum SURE as derived from
densely-sampled, high-resolution, non-accelerated data, and minimum SURE as derived from ACS are compared using
simulation experiments.
To avoid optimizing the rank of ESPIRiT's auto-calibrating matrix (one of the parameters), a heuristic derived from
SURE-based singular value thresholding is also proposed.

\noindent
\textbf{Results}: Simulations show SURE derived from the densely-sampled, high-resolution, non-accelerated data to be
an accurate estimator of the true mean squared error, enabling automatic parameter selection.
The parameters that minimize SURE as derived from ACS correspond well to the optimal parameters.
The soft-threshold heuristic improves computational efficiency while providing similar results to an exhaustive search.
In-vivo experiments verify the reliability of this method.

\noindent
\textbf{Conclusion}:
Using SURE to determine ESPIRiT parameters allows for automatic parameter selections.
In-vivo results are consistent with simulation and theoretical results.
  
\noindent
\textbf{Keywords}: Parallel Imaging Calibration, \emph{ESPIRiT}, \emph{Stein's Unbiased Risk Estimate}.

\clearpage

\section{Introduction}
\label{sec:introduction}

Modern MRI leverages multiple receive coil arrays to perform parallel imaging (PI) for faster acquisition-time and
improved signal to noise ratio (SNR).
The high level mechanics of PI can be broken up into two steps: calibration and reconstruction.
Calibration typically involves exploiting the data redundancy provided by the multiple receive channels to
derive linear data consistency operators that define the subspace the signal is expected to live in.
Reconstruction then enforces data consistency through these derived operators along with other priors to 
reconstruct the desired image.

Consider the two most common PI techniques, SENSE \cite{ref:sense} and GRAPPA \cite{ref:grappa}.
Both methods can be broken down into the two steps.
SENSE is an image-domain method that uses an explicit calibration scan to derive explicit spatial coil sensitivity maps (CSM),
which is then used as a spatial weighting operator during the reconstruction step.
GRAPPA uses auto-calibration signal (ACS) lines or a separate calibration scan to derive
convolution kernels in k-space that enforce data consistency.

In this work, the calibration of ESPIRiT \cite{ref:uecker} is studied.
ESPIRiT is a method that bridges SENSE and GRAPPA by using ACS to derive so called ESPIRiT maps
that can be used in a SENSE-like reconstruction.
In ESPIRiT, there are several user-set parameters in the calibration that determine the quality of the resulting ESPIRiT maps. 
These parameters are described in the theory section.
While ESPIRiT is in general robust to variation in these parameters, occasionally, poor parameter choice can result in 
reduced PI performance like noise amplification or image attenuation.
This is exemplified in Figure \ref{fig:variability}, where different parameter choices result in ESPIRiT maps with
significant variations in terms of signal attenuation and noise amplification.
This motivates the study of ESPIRiT parameter selection.

Since the quality of MRI reconstruction is often quantified in terms of Mean Squared Error (MSE), it is
desirable to select parameters that result in ESPIRiT maps that would result in a minimum MSE reconstruction.
However, calculating MSE as a function of the parameters is impossible without access to the true,
under-lying noise-free image.
To overcome this limitation, the parameter selection problem is reformulated into a denoising problem,
where ESPIRiT is viewed as a linear denoiser of the calibration data.
This formulation allows Stein's Unbiased Risk Estimate (SURE) \cite{ref:stein} to be used.
SURE is a technique that calculates the expected MSE of a given denoiser, and
has seen widespread applicability in the field of denoising \cite{luisier2007new, zhang1998adaptive,
blu2007sure, luisier2008sure}.
By calculating SURE as a function of ESPIRiT parameters, the expected MSE of the corresponding ESPIRiT
maps can be calculated and used to determine the optimal ESPIRiT map in this expected MSE sense.
Succinctly, SURE is used as a proxy to the underlying MSE for parameter optimization.

In order to calculate SURE for a given denoiser, it is required to calculate the divergence of the denoiser
with respect to the acquired data \cite{ref:stein, ref:ramani1}.
This is often very difficult for general denoising algorithms, resulting in the emergence of Monte-Carlo methods.
In particular, Ramani et al.\ demonstrates how Monte-Carlo simulations can calculate the divergence term required by
SURE for a general (possibly non-linear) denoiser and provides a black-box framework for denoiser parameter optimization
\cite{ref:ramani1}.
This Monte-Carlo SURE-framework has seen successful application to parallel imaging.
Weller et al.\ presents a SURE-based method of selecting regularization for k-space data-consistency linear
operators (GRAPPA \cite{ref:grappa} and SPIRiT \cite{lustig2010spirit}) that optimizes the linear operator
for reconstruction \cite{ref:weller1, ref:weller2}.

However, as discussed by Ramani et al., the divergence of a linear denoising operator corresponds to the trace of
the operator.
Consequently, if it is feasible to calculate the trace, the exact SURE value is obtained while avoiding Monte-Carlo
simulations.
This work demonstrates that, when considering non-accelerated, densely-sampled, high-resolution data, the exact SURE
value can be calculated as a function of ESPIRiT parameters.
In this case, ESPIRiT decouples into image-domain pixel-wise operators for which the divergence can be calculated exactly
to be the trace of said operators which can be calculated efficiently.
This work then augments ESPIRiT with a projection onto ACS to construct an ACS denoiser, and calculates the exact SURE
value for this approximate problem.
It is seen that the parameters corresponding to the minimum SURE as derived from ACS correspond
well to the optimal parameters derived from the minimum true squared error.

This allows for near-optimum parameter estimation while overcoming the need to perform Monte-Carlo simulations.
That being said, it is expected that the Monte-Carlo methods will achieve similar results with a larger computational cost.

This work calculates SURE for ESPIRiT by partitioning the complex vector space into real and imaginary parts, and
takes advantage of the Hermitian symmetric characteristics of the ESPIRiT operator for efficient calculation
of the divergence term required by SURE.
This is discussed in the Theory section and in the Appendix.
By partitioning the vector space, the application of SURE to ESPIRiT in this work fits within
the general linear denoiser framework presented by Ramani et al.\ \cite{ref:ramani1}.

Using SURE as a metric to quantitatively evaluate the performance of ESPIRiT parameters is desirable as it avoids adding
further model complexity and regularization to ESPIRiT.
This is in contrast to methods where regularization, additional iterations and other techniques are incorporated into the
model to build-in noise robustness while performing calibration \cite{xu2014robust, majumdar2013calibrationless, jin2016general,
park2012adaptive}.

While this work focuses on ESPIRiT, the main theme of this work is to demonstrate how the quantitative comparison of the denoising
efficacy of a PI operator can be used to perform optimal PI calibration.
The field of PI is extensive with many novel methods designed to fill different requirements and applications.
The proposed reformulation of the calibration step can be applied to many other novel PI methods as well.
For example, the matrix low-rank optimization in Parallel-LORAKS \cite{haldar2016p} can be reformulated into a matrix denoising problem,
where SURE can be used to find the optimal soft threshold to denoise the matrix \cite{ref:candes}.
The encoding matrix presented in Joint-SENSE \cite{ying2007joint} will also benefit from the application of SURE.
SURE can be calculated as a function of the polynomial coefficients used in Joint-SENSE to approximate CSM to determine which
derived CSM best denoises ACS data.
For GRAPPA based methods like iterative GRAPPA \cite{zhao2008iterative}, a Tikhonov parameter can be introduced into the least
squared fitting procedure, and this can be optimized over with SURE.
However, this reformulation is unlikely to benefit techniques involving non-linear functions acting on noise as the noise model
assumed by SURE no longer holds. For example, \cite{chang2012nonlinear, lyu2018kernl} involve non-linear mapping of data, and
\cite{majumdar2013calibrationless, jin2016general} explicitly rely on CS for noise robustness; both of which change the noise model
assumed by SURE.

Additionally, while this work presents the concept of ESPIRiT calibration based on its denoising efficacy, there have been numerous
works done on using SURE in the regime of regularization selection for MRI reconstruction.
For example, Ramani et al.\ presented a framework for tuning non-linear reconstructions based on their respective
Jacobians evaluated on acquired data \cite{ref:ramani2}; and Marin et al.\ presented a parameterized wavelet-based estimator that
uses SURE to determine the optimal parameters for reconstruction \cite{ref:marin}.
These works focus on selecting optimal regularization parameters for reconstruction, and often include a data-consistency
term in the reconstruction formulation.
Consequently, the SURE-based calibration in this work should work synergistically with these methods as it is expected to improve the
performance of the data-consistency operator.

Finally, one of the parameters in ESPIRiT determines the rank of a block Hankel structured auto-calibration matrix.
This work first demonstrates how SURE can be used to optimize for the same, then presents a singular value soft-thresholding
heuristic derived from the SURE-based low rank matrix approximation theory developed by Cand\`es et al.\ \cite{ref:candes}
to reduce computational overhead.
The heuristic uses the SURE-optimal soft threshold derived by Cand\`es et al. to weight the singular vectors of the auto-calibration
matrix.
This is seen to provide similar results to an exhaustive rank search while significantly reducing computational burden.
The details of this heuristic is discussed in the sections below.

It should be explicitly noted that this work assumes that noise is additive, complex normal noise that is not correlated.
If there is noise correlation between channels, the data should be whitened using noise characteristics derived
from either the data itself or an explicit noise acquisition.

\section{Theory}
\label{sec:theory}

\subsection{ESPIRiT}
ESPIRiT is a technique that combines the data-based calibration advantages of GRAPPA to derive 
SENSE-like relative CSM.
These CSM are derived from the null-space of the auto-calibration matrix, which is
formed by sweeping a kernel through the calibration region (as depicted in Figure \ref{fig:espirit}).
The following provides a brief overview of ESPIRiT, with emphasis on the parameters that 
need to be tuned.
For more detail about ESPIRiT and comparison to other PI methods, please refer
to \cite{ref:uecker}.

Let $A$ be the auto-calibration matrix, $V_{||}$ be a matrix consisting of the right singular
vectors of $A$ corresponding to the dominant singular values of $A$, $V_{\perp}$ be a matrix
consisting of the remaining singular vectors that span the null space of $A$, $U$ be a matrix
consisting of the left singular vectors of $A$ and $\Sigma$ be a diagonal matrix consisting of the
singular values of $A$ in descending magnitude order.
Then, by taking the singular value decomposition,
\begin{equation} 
A = U \Sigma V^* \text{ with } V = \mat{V_{||}, & V_{\perp}}.
\label{eq:svd} 
\end{equation}
Let $y$ be the noise-free, fully-sampled, underlying multi-channel k-space data (at the sampling locations).
Let $R_r$ be an operator that extracts a block from $y$ around the k-space position $r$
(including $r$ itself) across all the channels.
Since the signal $y$ should be orthogonal to the null-space of $A$, the following normal
equations are derived:
\begin{subequations} 
    \begin{equation} 
    \left(\sum_r R_r^* V_\perp V_\perp^* R_r\right) y = 0
    \end{equation}
    \begin{equation} 
    \left(\sum_r R_r^* \left(I - V_{||} V_{||}^* \right)R_r\right) y = 0
    \end{equation}
    \begin{equation}
    \underbrace{\left(\sum_r R_r^* R_r\right)^{-1}
    \left(\sum_r R_r^* \left(V_{||} V_{||}^* \right)R_r\right)}_{\mathcal{W}} y = y 
    \label{eq:espiritconst}
    \end{equation}
\label{eq:espiritnormal}
\end{subequations}
$\left(\sum_r R_r^* R_r\right)^{-1}$ effectively scales each channel by a scalar that is the 
inverse of the number of k-space elements selected by $R_r$.
Thus, $\mathcal{W}$ is a convolution with a matrix-valued kernel where the matrix operates
on the channel dimension.
This convolution operator is decoupled into pixel-wise operations along the channel dimension in
the image domain.
In other words, let $x$ be the true multi-channel image data such that $y = F x$ where $F$ is
the unitary Fourier transform operator.
Equation \eqref{eq:espiritconst} becomes,
\begin{equation}
\underbrace{(F^* \mathcal{W} F)}_{\mathcal{G}} x = x,
\label{eq:espiritimconst}
\end{equation}
where $\mathcal{G}$ is defined to be $F^*\mathcal{W}F$.
Particularly, since $\mathcal{G}$ is decoupled into pixel-wise operators, it suffices to look at
the effect of $\mathcal{G}$ on a particular image pixel with index $q$.
\begin{equation}
\mathcal{G}(q) x(q) = x(q) 
\label{eq:espiritimconstpix}
\end{equation}
$x(q)$ is a vector of dimension equal to the number of CSM.
Let the source image be $m$ and let $S$ be the vector constructed from stacking the coil 
sensitivities of the different channels.
The SENSE model states,
\begin{equation}
x(q) = S(q) m(q) 
\label{eq:forwardmodel}
\end{equation}
$m(q)$ is a scalar and $S(q)$ is a vector of dimension equal to the number of coil-sensitivity
maps.
Applying this to \eqref{eq:espiritimconstpix},
\begin{equation}
\mathcal{G}(q) S(q) m(q) = S(q) m(q)
\end{equation}
If $m(q) \neq 0$, the following eigenvalue-eigenvector condition is derived:
\begin{equation}
\mathcal{G}(q) S(q) = 1 \cdot S(q)
\label{eq:espiriteig}
\end{equation}
Thus sensitivity maps, in the ideal case, are eigenvectors of $\mathcal{G}$ with eigenvalues one,
with the other eigenvectors of $\mathcal{G}$ having eigenvalues much smaller than one.
This comes from observing that $\mathcal{W}$ is an average of projections and is 
consequently positive semi-definite with eigenvalues smaller or equal to one.
In practice, due to data-inconsistencies (like noise), the observed eigenvalues of the eigenvector
maps are very close to (but not exactly) one.
This motivates defining an approximate ``$\approx 1$'' condition where the eigenvalues that would be
$1$ in the ideal case but are instead close to one.
The eigenvalue decomposition of $\mathcal{G}$ is taken and the eigenvector corresponding to eigenvalue
``$\approx 1$'' are considered to be ESPIRiT maps.

In some cases, multiple eigenvalues ``$\approx 1$'' appear such as
when the calibration region supports a field-of-view smaller than the object which results 
in multiple sensitivity values at a pixel location due to aliasing \cite{ref:uecker}.
This motivates using more that one set of these eigenvector maps to better capture the desired signal.
\subsection{Parameter choices in ESPIRiT} 
There are three parameters in the ESPIRiT calibration.
The first is the size of the window, or kernel, that is swept through the ACS to construct the
auto-calibration matrix $A$; the second is the size of the signal subspace used to partition $V_{||}$ from
$V_{\perp}$; and the third is the threshold, above which eigenvalues are considered ``$\approx 1$''.
These parameters are denoted as the kernel size $(k)$, the subspace size $(w)$ and the eigenvalue crop threshold
$(c)$, respectively.

Let $n_c$ be the number of channels.
For 2D data, a kernel size $(k)$ would imply that window of dimensions $(k \times k \times n_c)$ is swept
through the calibration region to construct the rows of the auto-calibration matrix $A$.
Consequently, $V_{||}$ and $V_{\perp}$ together span a linear space of dimension
$(k \cdot k \cdot n_c)$, and the rank of $V_{||}$ can vary from $0$ to $(k \cdot k \cdot n_c)$.
The chosen rank of $V_{||}$ is the subspace size $(w)$ and is measured in Window Normalized Singular 
Values Number (WNSVN).
This normalizes the rank by the kernel spatial dimensions and thus $(w)$ is in the range from $0$ to $n_c$.
A subspace size of $(w)$ implies $V_{||}$ consists of $w \cdot k^2$ orthogonal vectors.
Finally, the eigenvalue crop threshold $(c)$ determines the pixel positions $(q)$ 
where the eigenvectors of operator $\mathcal{G}(q)$ are well defined, which corresponds to pixels positions
$(q)$ within the object.
A too high threshold would result in direct attenuation of the signal, and too small a threshold would allow
eigenvectors of operator $\mathcal{G}(q)$ from positions that are not well defined in terms of
\eqref{eq:espiriteig}, which in turn allows in signal that is not necessarily from the object, such as noise.
The feasible values of the eigenvalue crop threshold $(c)$ are from $0$ to $1$, with realistic values residing
in the range from $0.7$ to $0.95$.
The rank of $V_{||}$ also represents a similar trade-off.
Since the same under-lying data is observed through multiple receive channels, the auto-calibration matrix $A$ is 
expected to be low-rank, but this is often not the case due to noise and other data inconsistencies.
Since the ESPIRiT operator is derived from $V_{||}$, too large a subspace size implies that the ESPIRiT operator
captures these inconsistencies within its range space while too small a subspace size yields an operator with 
insufficient information to properly describe the underlying signal.
In the latter case, a projection onto the operator's range space will result in some signal loss.
Lastly, the kernel size $(k)$ of ESPIRiT determines the smoothness of the resulting ESPIRiT maps, where a smaller
kernel results in smoother CSM.
However, a too small kernel size may result in a calibration matrix $A$ whose $V_{||}$ does not
contain sufficient information to create the SENSE-like operator.

While ESPIRiT is fairly robust to these parameter choices, there is variability in map 
quality (such as how well the maps capture the field of view of the object) and 
choosing parameters that result in optimal reconstruction is desirable.
Figure \ref{fig:variability} exemplifies the variability in ESPIRiT maps when varying the subspace 
size $(w)$ and eigenvalue crop threshold $(c)$ for a fixed kernel size $(k)$. 

In order to develop a robust, data-driven method of automatic parameter selection, 
Stein's unbiased risk estimate (SURE) is explored as a metric to select parameters 
that are optimal in an expected mean squared error sense.
\subsection{Stein's unbiased risk estimate}
Stein's unbiased risk estimate is a data-driven method of calculating the expected mean
squared error of an estimator in the presence of zero-mean additive Gaussian noise, given that
the estimator is differentiable with respect to the data almost everywhere \cite{ref:stein}.
The SURE expression for a Hermitian symmetric operator is presented and then extended to ESPIRiT.

Let $x \in \mathbb{C}^m$ be the ground truth to be estimated; $n$ be zero-mean, additive, 
Gaussian complex noise with standard deviation $\sigma$; and $x_\text{acq} = x + n$ be the 
acquired data.
Let $\pmb{P}_\theta \in \mathbb{C}^{m\times m}$ be a Hermitian symmetric linear operator,
parameterized by $\theta$, that is an estimator of $x$ for $x_{\text{acq}}$.

This work assumes that the noise is additive, normal noise that is not correlated.
If there is noise correlation between channels, the data should be pre-whitened using noise
characteristics derived either from the data itself or from an explicit noise measurement.

Let $\E{\cdot}$ denote the expected value operation.
Partitioning the complex vector space into real and imaginary parts and noting that the divergence
of a linear operator is the trace of the linear operator (steps described in the appendix), Stein's
first theorem \cite{ref:stein} implies:
\begin{subequations}
  \begin{equation}
  \E{\norm{\pmb{P}_\theta x_\text{acq} - x}_2^2} = \E{SURE_{\pmb{P}_\theta} (x_\text{acq})}
  \label{eq:surebasic_a} 
  \end{equation}
  \begin{equation} \begin{array}{rcl} SURE_{\pmb{P}_\theta} \left(x_\text{acq}\right) 
    &=& -m \sigma^2 + \norm{\left(\pmb{P}_\theta - I\right)x_\text{acq}}_2^2 + 2 \sigma^2 \,
        \left[\text{div}_{x_\text{acq}}\left(\pmb{P}_\theta\right)\right]\left(x_\text{acq}\right) \\
    &=& -m \sigma^2 + \norm{\left(\pmb{P}_\theta - I\right)x_\text{acq}}_2^2 + 2 \sigma^2 \text{trace}
        \left(\pmb{P}_\theta\right)
  \end{array} \label{eq:surebasic_b} 
  \end{equation}
  \label{eq:surebasic}
\end{subequations}
Particularly, note that Equation $\eqref{eq:surebasic_b}$ is independent of $x$, and only 
depends on the acquired data $x_\text{acq}$.
Thus SURE can be used as a surrogate for the expected mean squared error to find the optimal
parameters $\theta^*$.
\subsection{SURE with ESPIRiT}
The main concepts will first be illustrated through a non-accelerated, densely-sampled,
high-resolution case where the requirement of zero-mean additive normal noise is satisfied.
In this case, there will be no noise-amplification due to data interpolation and the efficacy of the ESPIRiT
operator is tied to the quality of the denoising it performs.
This will later extend to the case when the densely-sampled region is restricted to the low-resolution
ACS signal.

\begin{figure}[ht] \centering
\includegraphics[width=0.95\textwidth]{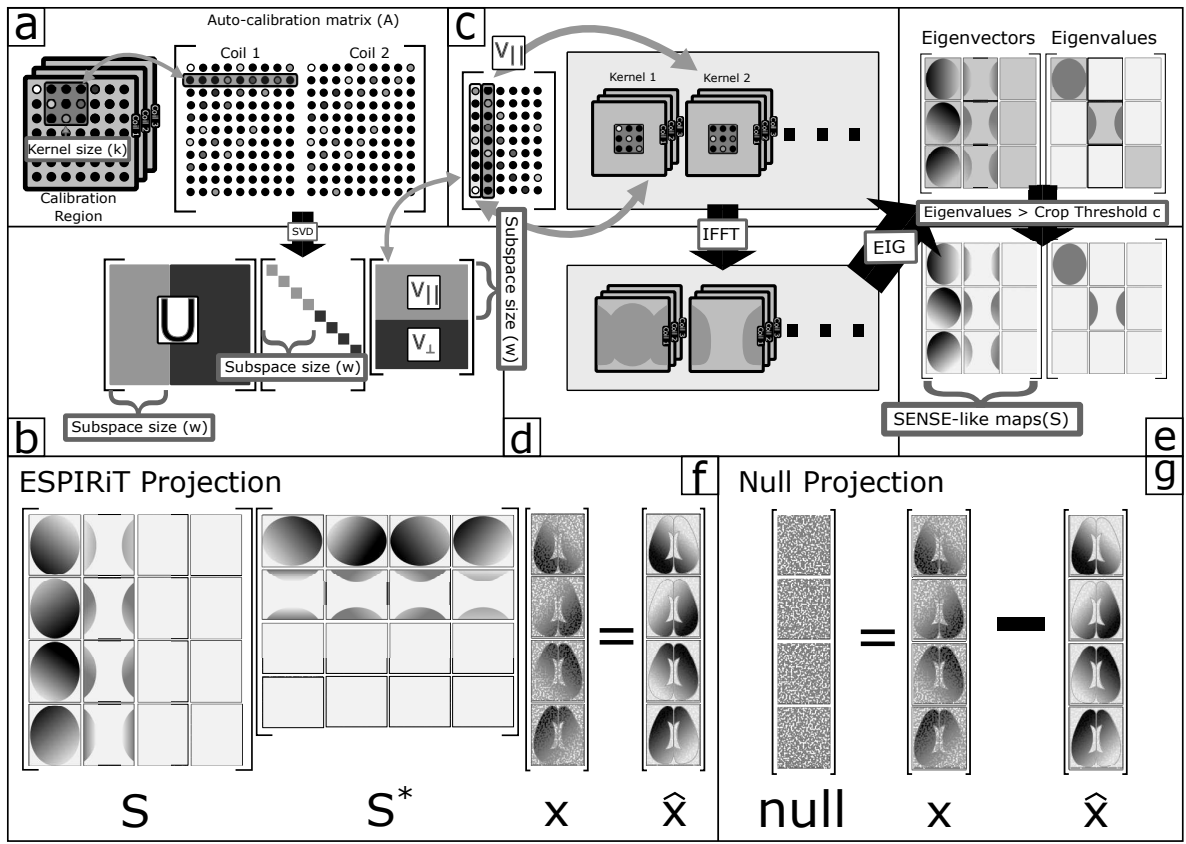}
\caption{
  This summarizes the technique of generating ESPIRiT maps that are used as 
  SENSE-like projection operators. (a) A kernel is swept through the calibration region to
  construct the calibration matrix (which consequently has block Hankel structure). (b) The
  SVD of the calibration matrix is taken and (c) the right singular vectors corresponding to the
  largest singular values are reshaped into k-space kernels. (d) The inverse Fourier transform
  of these kernels are taken, followed by (e) the eigenvalue decomposition of each pixel along the
  coil dimension in the image domain. The eigenvectors corresponding to eigenvalues 
  ``$\approx 1$" are used to construct the ESPIRiT operator. (f) depicts how ESPIRiT maps
  can be used as a projection operator to denoise data in the ideal case. (g) Consequently, 
  the null space of the ideal projection operator would contain only noise.
}
\label{fig:espirit}
\end{figure}

To quantify denoising, first define an ESPIRiT projection operator, which denoises the acquired data using
a projection onto the ESPIRiT maps.
An illustration is provided in Figure \ref{fig:espirit}f.

Let $x_{acq}$ denote the acquired multi-channel images obtained from applying a Discrete Fourier transform
on the non-accelerated, densely-sampled, high-resolution acquired k-space.
Let $n_c$ be the number of coils.
Let $S^i(q)$ be the $i^{th}$ eigenvector of $\mathcal{G}(q)$ with eigenvalue $\lambda_i(q)$.
$S^i(q)$ is a vector of dimension $n_c$ and has unit norm.
Since $\mathcal{G}$, and consequently $\mathcal{G}(q)$, is Hermitian symmetric, the eigenvectors $S^i(q)$ will be orthonormal to each other.
The ESPIRiT projection operator at pixel position $q$, which will be denoted as $P(q)$, is defined as:
\begin{equation}
\pmb{P}(q) = 
\mat{| & \null & | \\ S^1(q) & \dots & S^{n_c}(q) \\ | & \null & |}
\mat{| & \null & | \\ S^1(q) & \dots & S^{n_c}(q) \\ | & \null & |}^*
\label{eq:Scomp}
\end{equation}
The aggregate ESPIRiT projection operator $\pmb{P}$ can be represented by stacking the pixel-wise 
operators diagonally:
\begin{equation}
\pmb{P} = \mat{
  \pmb{P}(1) & \hdots & 0      \\
  \vdots     & \ddots & \vdots \\
  0          & \hdots & \pmb{P}(N)
} \text{ where $N$ is the number of image pixels.}
\label{eq:espiritprojop}
\end{equation}
The range space of the ESPIRiT projection operator describes the linear subspace the desired signal is
expected to reside in.
Projecting onto this subspace is expected to remove undesirable signal such as additive,
white, Gaussian noise that contaminates MRI data.
This interpretation of ESPIRiT as a denoiser allows for the application of SURE.

Let $\pmb{P}_\theta$ be the projection operator derived from a particular ESPIRiT parameter set 
$\theta = (k, w, c)$.
Define $x$ as the densely-sampled, high-resolution, noise-less, multi-channel images; $n$ as additive,
complex normal noise of standard deviation $\sigma$; $x_\text{acq} = x + n$ as the acquired coil images; and $I$
as the identity operator.
In this densely-sampled, high-resolution case, finding the optimal projection operator is reduced to finding
the optimum $\theta^*$ that results in a projection operator $\pmb{P}_{\theta^*}$ that best denoises the input 
data $x_\text{acq}$.
That is to say, 
\begin{equation}
\theta^* = \argmin_\theta \norm{x - \pmb{P}_\theta x_\text{acq}}_2^2
\label{eq:opt}
\end{equation}
Since ESPIRiT is a pixel-wise linear projection operator in the image domain, the divergence
contributed by a single pixel $x(q)$ is the trace of the linear operator affecting that pixel 
(denoted $\pmb{P}_\theta(q)$) (described in the appendix).
Thus, the SURE value of $\pmb{P}_\theta$ can be calculated by summing over all pixel positions $q$. 
\begin{equation} 
SURE_\theta(v) = \sum_q \left[-n_c \sigma^2 + \norm{\left(\pmb{P}_\theta(q) - 
I\right)x_\text{acq}(q)}_2^2 + 2 \sigma^2 \text{trace } \pmb{P}_\theta(q)\right] 
\label{eq:sureespirit}
\end{equation}
Equation $\eqref{eq:sureespirit}$ can then be used as a surrogate for $\eqref{eq:opt}$.
\begin{equation}
\theta^* = \argmin_\theta \norm{\pmb{P}_\theta x_\text{acq} - x}_2^2 \approx \argmin_\theta 
SURE_\theta\left(x_\text{acq}\right)
\label{eq:suresol}
\end{equation}
Since the trace of $AB$ is equal to the trace of $BA$ for any matrix $A$ and $B$, and since $S^i(q)$ are
orthonormal to each other, the trace of $\pmb{P}_\theta$ can be efficiently calculated by permuting the
matrices in Equation \eqref{eq:Scomp}.
\begin{equation}
\text{trace } \pmb{P}_\theta = \sum_q \sum_i \norm{S^i(q)}^2
\label{eq:trace}
\end{equation}
Thus, in the above case, it is possible to search through values of $\theta$ to determine the optimal
projection operator $\pmb{P}_\theta$ for calibration.
Equation \eqref{eq:suresol} is used as a basis to optimize for the ESPIRiT projection operator.
A variant of \eqref{eq:suresol} is presented in the following for the case in which the densely sampled
region is limited to only the ACS data.
\subsection{Accelerated case with the auto-calibration signal}
In order to accelerate PI scans while still being auto-calibrating, the ESPIRiT operator is
estimated from only the low-resolution, densely sampled ACS data.
This is accomplished by incorporating a Fourier sampling operator into the projection operator $\pmb{P}$ and
enforcing data consistency within the ACS data in k-space.

Define $R$ to be the operator that outputs an ACS region from densely-sampled k-space. 
Let a new, augmented projection operator $\pmb{P}_\theta^R$ be defined as follows:
\begin{equation}
\pmb{P}_\theta^R = R F S S^* F^* R^*
\label{eq:espiritprojopcal}
\end{equation} 
Let $y$ be the noise-less, densely-sampled, low-resolution auto-calibration region in k-space; 
$n$ be additive, zero-mean, complex normal noise of standard deviation $\sigma$; and 
$y_\text{acq} = y + n$ be the acquired auto-calibration region.
Then, interpreting \eqref{eq:espiritprojopcal} as a denoiser of ACS data yields the following SURE expression:
\begin{equation} 
SURE_\theta(y)= \norm{\left(\pmb{P}_\theta^R-I\right) y_\text{acq}}_2^2 + 2 \sigma^2 
\text{trace } \pmb{P}_\theta^R + C 
\label{eq:sureespiritcal}
\end{equation}
$C$ is some constant term that is ignored because it does not affect the minimum.
Since the trace of a matrix does not depend on a basis and $F^*$ is a change of basis, it follows that,
\begin{equation}
\text{trace } \pmb{P}_\theta^R = \text{trace } R F S S^* F^* R^* = \text{trace } F^* R F S S^* F^* R^* F
\end{equation}
Thus, the trace can be calculated efficiently in a manner similar to Equation \eqref{eq:trace}.
\begin{equation}
\text{trace } \pmb{P}_\theta^R = \sum_q \sum_i \norm{\underline{S}^i(q)}^2 \text{ where } \underline{S} = F^*RFS
\end{equation}
Note that the exact SURE value is calculated for a denoiser constructed by augmenting ESPIRiT with a projection onto ACS.
Since this is a different denoising operator (compared to the previous densely-sampled, high-resolution non-accelerated
case), the SURE values calculated (via Equations $\eqref{eq:sureespiritcal}$ and $\eqref{eq:sureespirit}$) will not necessarily
match.
That being said, the SURE values calculated by Equations $\eqref{eq:sureespiritcal}$ and $\eqref{eq:sureespirit}$ are exact for their
respective denoisers.

The parameters $(\theta)$ obtained from minimizing Equations \eqref{eq:sureespirit} and \eqref{eq:sureespiritcal}
often correspond in practice.
With the assumption that the optimal ESPIRiT maps are derived from parameters that best denoise densely-sampled,
high-resolution, non-accelerated data, near optimal parameter selection while being limited to ACS is achieved.

Enforcing consistency of ACS data and the corresponding SURE expression is seen to be a good representative of the
performance of ESPIRiT maps.
This allows different parameter sets to be searched through to obtain the ESPIRiT projection operator that results in
near-optimal performance in the expected mean squared error sense even when restricted to ACS data.
\subsection{Soft-threshold based weighting for subspace estimation}
In practice, sweeping through different kernel sizes and signal subspace sizes (or the rank 
of $A$ and $V_{||}$) is computationally intensive whereas sweeping through different thresholds to 
determine the eigenvalue crop threshold $(c)$ is relatively quick.
To aid viability, a heuristic that appropriately weights the right singular vectors based on singular value 
soft-thresholding as an alternative to sweeping through different rank values is presented.

Since the same underlying data is being observed through multiple channels, the auto-calibration matrix 
$A$ is expected to be low rank.
However, due to noise and other data inconsistencies present in the data, the observed auto-calibration
matrix often has full rank.
A low-rank matrix estimate of $A$ is then constructed by hard thresholding the singular values of $A$.
Ideally, it would be beneficial to use SURE to determine the optimal low-rank matrix estimate in the sense of
``denoising'' the matrix.
However, this is difficult to do since a hard threshold is not weakly-differentiable.
A common alternative is to soft-threshold the singular values.

Consider the singular value decomposition of $A$ in its dyadic form.
\begin{equation} 
A = \sum_i s_i u_i v_i^* 
\end{equation}
Here, $u_i$ are the left singular vectors, $v_i$ are the right singular vectors and $s_i$ are 
the singular values.
A soft-threshold low rank matrix estimate of $A$, which will be denoted as $\widehat{A}$, is constructed by 
soft-thresholding the singular values by threshold $\lambda$.
\begin{equation} 
\widehat{A} = \sum_i (s_i-\lambda)_+ u_i v_i^* 
\label{eq:lowrankest}
\end{equation}

If $A$ does not contain structured noise, Cand\`es et al.\ showed that it is possible to use SURE to efficiently
find the optimal $\lambda$ in Equation \eqref{eq:lowrankest}, which will be denoted as $\lambda^*$ \cite{ref:candes}.
However, since $A$ is block-Hankel, it contains structured noise.
That being said, in this work, the method derived by Cand\`es et al. is applied directly to obtain an approximate $\lambda^*$.
With more relaxed computational efficiency requirements, black-box Monte-Carlo methods may be utilized to estimate $\lambda^*$, such
as the work by Ramani et al.\ \cite{ref:ramani1}.
If computational efficiency is a non-issue, it is possible to utilize Equations \eqref{eq:sureespirit} and \eqref{eq:sureespiritcal}
to enumerate all possible ranks of matrix $A$ to determine the optimal rank of $A$.
This is demonstrated by Experiment (a) in the Theory section.

The goal of this heuristic is to avoid enumerating the rank of $A$ by incorporating $\lambda^*$ to form a weighted subspace estimate.
To motivate weighting the singular vectors, consider the following.
The subspace selection problem can be modeled by a hard threshold on the singular values using a
threshold $\lambda$.
\begin{equation} 
V_{||} = V W \text{ where } W \text{ is a diagonal weight matrix with } 
W_{ii} = \left\{\begin{array}{rl}1, & \left(s_i > \lambda\right)\\
                                 0, & \text{otherwise}\end{array}\right.
\label{eq:surehardthresh}
\end{equation}
Instead, weight the singular vectors with its soft-threshold variant that is used in
\eqref{eq:lowrankest}.
Let this weighted subspace be $V_{||}^w$.
\begin{equation} 
V_{||}^w = V W \text{ where } W \text{ is a diagonal weight matrix with } 
W_{ii} = \frac{\left(s_i - \lambda \right)_+}{s_i}
\label{eq:sureweightsub}
\end{equation}
The parameter $\lambda^*$ is calculated by Cand\`es' SURE method in \eqref{eq:sureweightsub} to get a 
weighted subspace estimate $V_{||}^w$. This weighted subspace estimate is then used in Equation \eqref{eq:espiritconst} instead of $V_{||}$.

The ESPIRiT operators derived from $V_{||}^w$ retain the differentiability property utilized by Equations \eqref{eq:sureespirit} and
\eqref{eq:sureespiritcal}.
Consequently, the exact SURE value as a function of the crop threshold can be calculated for ESPIRiT operators derived from $V_{||}^w$.

\section{Methods}
\label{sec:methods}

To verify the efficacy of the technique, exhaustive simulation experiments are conducted using MATLAB (MathWorks, Natick, MA)
to compare the true squared error and the expected risk as calculated by SURE for ESPIRiT.
The feasibility of the method is demonstrated by conducting in-vivo experiments using ESPIRiT.
Additionally, the soft-threshold heuristic with an eigenvalue crop threshold search (using a method akin to line-search) is implemented
in the Berkeley Advanced Reconstruction Toolbox (BART) \cite{ref:uecker2}.
\subsection{Simulation Experiments}
Fully-sampled, high-resolution data of the human brain were acquired on a 1.5T scanner (GE, Waukesha, WI) using 
an eight-channel coil for a subject (with IRB approval and informed consent obtained).
Data were collected obtained using inversion-recovery prepared 3D RF-spoiled gradient-echo sequence with the following
parameters: $TR/TE = 12.2/5.2\;ms$, $TI = 450 ms$, $FA = 20^\circ$, $BW = 15kHz$, and a matrix size of 
$256\times180\times230$ with $1\; mm$ isotropic resolution.
This was done to have a ground-truth to compare against and verify the accuracy of SURE as an estimator of the
mean squared error. The $3D$ dataset was Fourier transformed along the readout direction and a slice along the
readout dimension was taken.
ESPIRiT maps were calculated from this slice and the projection of this slice onto the ESPIRiT maps was 
considered to be the true, underlying ground truth.
The ground truth has dimensions $230 \times 180 \times 8$, where $8$ is the number of channels, and an
$l_2-$norm of $8369.46$.
Additive complex k-space noise of standard deviation $4$ was retrospectively added to the ground-truth
and the result was considered to be the acquired data.

For different parameter values $(\theta)$, ESPIRiT maps were generated and the true squared
error between the projection and the ground truth was calculated.
This error was compared to the calculated SURE value.
The soft-threshold based weighting heuristic was also tested by similarly varying kernel sizes and eigenvalue
crop thresholds and comparing the true squared error to SURE.
For all the previous cases, the SURE approximation when being restricted to low-resolution, densely-sampled
ACS data was also calculated.

In Experiment (a), a fixed kernel size of $6$ was used and the subspace size and eigenvalue crop 
thresholds were varied.
For each subspace size and eigenvalue crop threshold, the true squared error, SURE value given high-resolution, 
densely sampled data, and the SURE value given low-resolution, densely-sampled ACS data were calculated.
The latter curve was normalized by a constant to better compare the minimums of each of the curves.

In Experiment (b), a fixed kernel size of $6$ was used along with the soft-threshold based 
subspace weighting heuristic.
The true squared error, SURE value given high-resolution, densely sampled data, and the SURE value given low-resolution,
densely-sampled ACS data was calculated as the crop threshold was varied.
The latter curve was normalized by a constant to better compare the minimum of each of the curves.

In Experiment (c), the same three curves were calculated from varying the kernel size, subspace 
size and eigenvalue crop threshold parameters.
The minima given a particular kernel size across the subspace thresholds and eigenvalue crop thresholds for
that kernel size were taken.
This experiment was conducted to test the dependence of ESPIRiT maps on the kernel size $(k)$ assuming optimal 
subspace size and eigenvalue crop threshold for that particular kernel size $(k)$.

In Experiment (d), the same three curves were calculated from varying the kernel size and eigenvalue crop 
threshold parameters while using the soft-threshold based subspace weighting heuristic.
The minima across the eigenvalue crop thresholds given a particular kernel size were taken.
This experiment was conducted to test the dependence of ESPIRiT maps on the kernel size $(k)$ assuming an
optimal eigenvalue crop threshold for that particular kernel size $(k)$.

In Experiment (e), a fixed kernel size of $6$ was used along with the soft-threshold based subspace weighting 
heuristic, and the eigenvalue crop threshold was varied.
For each eigenvalue crop threshold, the true squared error and corresponding g-factor maps are calculated.
The maximum and average g-factor within the field of view of the desired object is used to demonstrate
how ESPIRiT maps calibrated according to the minimum squared error can aid with accelerated acquisitions.
The g-factor maps are calculated assuming equispaced $2\times 2$ sub-sampling in the phase encode directions.

Experiments (a, b, c, d, e) used an ACS size of $24 \times 24$.
\subsection{In-Vivo Experiments}

In Experiment (f), a 3D accelerated dataset was acquired on a 3T Achieva scanner (Philips,
Best, The Netherlands) with IRB approval and informed consent obtained.
The acquisition was a $T_1$-weighted, TFE dataset acquired using Poisson disk under-sampling ($R \sim 2$) with $18\times18$
ACS lines using an 8-channel head coil.
The k-space data were pre-whitened using scanner software based on noise measurement. 
A unitary inverse Fourier transform was taken along the readout direction and a slice was extracted.
ESPIRiT calibration was performed on the dataset using parameters selected by the SURE-based method with
the soft-threshold based weighting heuristic.
A fixed kernel size of $6$ is used along with the line-search like method for calculating the eigenvalue
crop threshold.
A PI+compressed sensing (CS) iterative reconstruction was performed with the SURE-calibrated ESPIRiT maps
(with an $l_1$ regularization of $0.01$ using a wavelet sparsity prior). 
The iterative reconstruction used BART. 

In Experiment (g), one pre-whitened 3D dataset was acquired on a 3T Skyra scanner (Siemens Healthcare,
Erlangen, Germany) with IRB approval and informed consent obtained.
The acquisition was a high resolution, densely sampled $T_2$-weighted, TSE dataset with no under-sampling.
The data is coil compressed from 32 channels to 8 channels using geometric coil compression \cite{zhang2013coil}.
A unitary inverse Fourier transform was taken along the readout direction and a slice was extracted.
The densely sampled data was retrospectively under-sampled using a $2 \times 2$ Poisson disk sampling mask and a 
$2 \times 2$ equispaced sampling mask.
ESPIRiT calibration was performed on the dataset using parameters selected by the SURE-based method with
the soft-threshold based weighting heuristic.
A fixed kernel size of $6$ is used along with the line-search like method for calculating the eigenvalue
crop threshold.
PI+CS iterative reconstructions was performed with the SURE-calibrated ESPIRiT maps
(with an $l_1$ regularization of $0.01$ using a wavelet sparsity prior). 
These iterative reconstructions were implemented in BART. 

In Experiment (h), the SURE-based parameter selection is applied to the same data used in the original ESPIRiT
work \cite{ref:uecker} that demonstrated ESPIRiT's robustness to aliasing due to the calibration region supporting
a FOV smaller than the object.
The same 2D spin-echo dataset ($TR/TE=550/14 ms$, $FA=90^\circ$, $BW = 19 kHz$, matrix size: $320\times168$,
slice thickness: $3 mm$, 24 reference lines) with an FOV of $(200 \times 150) mm^2$, acquired at $1.5T$ using
an $8$-channel head coil is used.
Additionally, the data is retrospectively under-sampled using an equi-spaced $2\times2$ sampling mask with
an ACS size of $24 \times 24$.
The data was Fourier transformed to the image domain and noise variance was estimated from a corner of the image 
data that did not contain any desired signal.
To determine the ESPIRiT maps, a fixed kernel size of $6$ is used along with the soft-threshold based weighted 
subspace estimate. The eigenvalue crop threshold was calculated using the line-search like method.
The resulting ESPIRiT map was then calculated to verify robustness to aliasing.
\subsection{Characterization Experiments}

Experiment (i): Using the same data as in Experiment (a), ESPIRiT maps are calculated
as a function of calibration size $(r)$ using the soft-weighting heuristic and MSE-optimal crop threshold.
A calibration size of $r$ implies $(r \times r)$ ACS lines.
The resulting projection MSE is also calculated.
This is to characterize the effect of the number of ACS lines in deriving optimal ESPIRiT maps.

Experiment (j) explores the possibility of using SURE for more finely-tuned parameterization by calibrating
ESPIRiT maps on a slice-by-slice basis that is SURE-optimal per slice.
This should allow for ESPIRiT's parameters to vary as a function of the noise level per slice.
This concept is applied to the $T_2$-weighted dataset in Experiment (g), where the SURE-calibration
uses the soft-weighting heuristic.
Additionally, this experiment compares the SURE-optimal calibration to default calibration parameters present
currently in BART.

Experiment (k): It is also worthwhile considering the effects of kernel size when using the soft-threshold heuristic
and the SURE-optimized crop threshold.
ESPIRiT calibration is performed on the same dataset as in Experiment (a), and g-factor is
calculated as a function of kernel size $(k)$.

\section{Results}
\label{sec:results}

\subsection{Simulation Results}
The simulation results of Experiments (a), (b), (c) and (d) are illustrated in Figure \ref{fig:espirit_sim}.
In Experiments (a) and (b), it is seen that the true squared error calculated with the densely-sampled, high-resolution
data and SURE correspond well.
It is also seen that SURE as calculated from ACS data has a minimum close to the minimum of the true
squared error.
Additionally, the minimum true squared error of Experiment (b) is higher than the minimum true squared error of
Experiment (a) by approximately $1.34\%$, demonstrating that the soft-threshold heuristic yields ESPIRiT maps that perform
almost as well as the optimal EPSIRiT maps derived from enumerating through the rank of the auto-calibration matrix.
With comparable MATLAB (MathWorks, Natick, MA) implementations, the exhaustive rank search and crop threshold search in
Experiment (a) took days to complete while the exhaustive crop threshold search with the soft-threshold heuristic in Experiment (b)
completed in a few hours on an Intel(R) Xeon(R) Gold 6138 CPU.
In comparison, for Experiment (h), the soft-threshold heuristic and crop threshold search implemented in BART completes in $7.112$
seconds on the same CPU.
Thus significant computational gain is achieved with minimal performance penalty.
Experiments (c) and (d) do not show perfect correspondence.
However, the squared error in these cases are in the order of $10^5$, which is much smaller than Experiments (a) and
(b), where the squared errors and in the order of $10^7$.

The simulation results of Experiment (e) is illustrated in Figure \ref{fig:gfact}.
Note how until Point B, a decrease in the squared error corresponds to better g-factor performance,
while any increase in the crop threshold beyond Point B attenuates the signal.
This experiment demonstrates that, even though a crop threshold value reflect in Point A is sufficient,
there is g-factor performance to be gained through optimizing for the minimum squared error.

On simulated data that fits the model, it is seen that SURE as estimated from densely-sampled, high-resolution non-accelerated
data is an accurate estimator of the squared error.
Furthermore, restricting ourselves to ACS results in near-optimal parameter choice.
\subsection{In-Vivo Results}
\begin{figure}[ht] \centering
\includegraphics[width=0.95\textwidth]{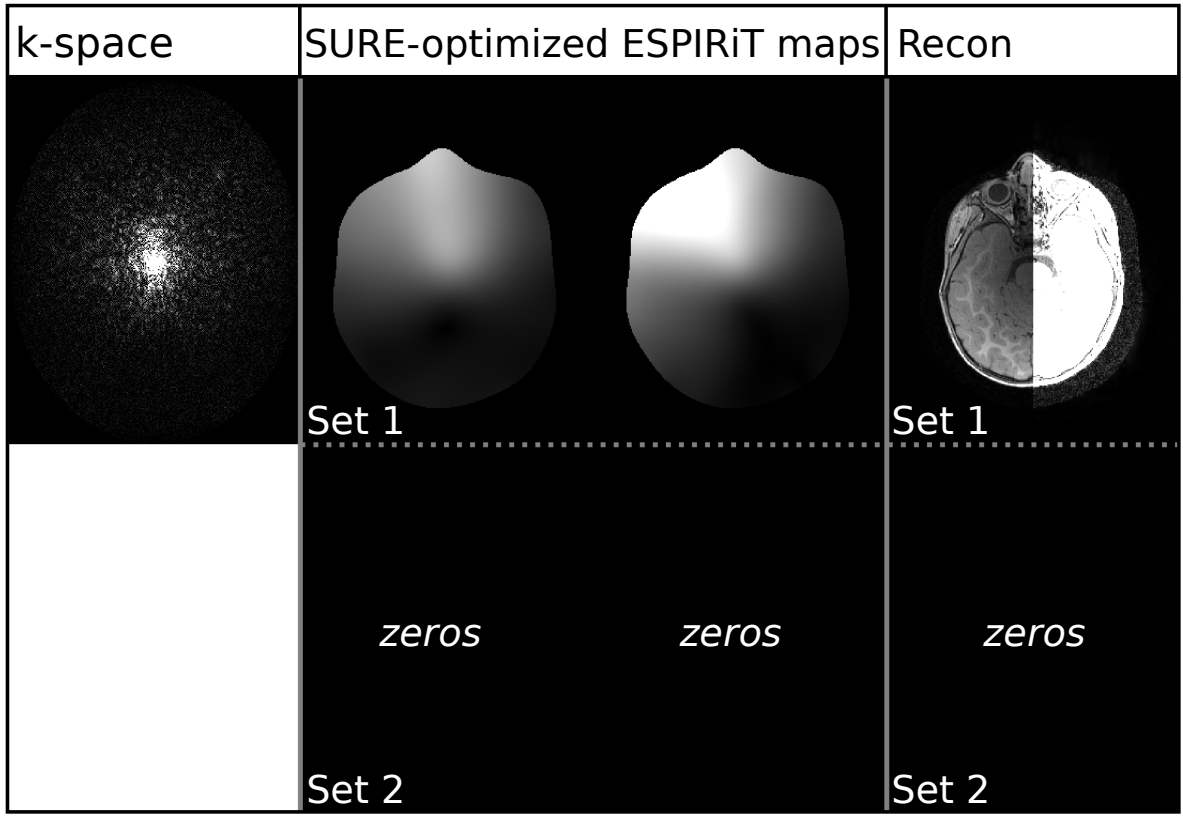}
\caption{
  Experiment (f):
  This figure demonstrates the application of SURE to calibrate ESPIRiT maps on a $T_1$-weighted acquisition
  where roughly $50\%$ of the total phase encode points are acquired, which included an ACS size of $18\times 18$.
  Half of the image's contrast has been enhanced to illustrate how well the field of view of the desired object
  has been captured.
  ``Set 1" and ``Set 2" refers the ``soft" SENSE model, where multiple ESPIRiT maps may satisfy the ``$\approx 1$"
  eigenvalue condition.
  $2$ of the $8$ calibrated ESPIRiT maps are depicted in this figure for brevity.
}
\label{fig:invivo_t1}
\end{figure}

\begin{figure}[ht] \centering
\includegraphics[width=0.75\textwidth]{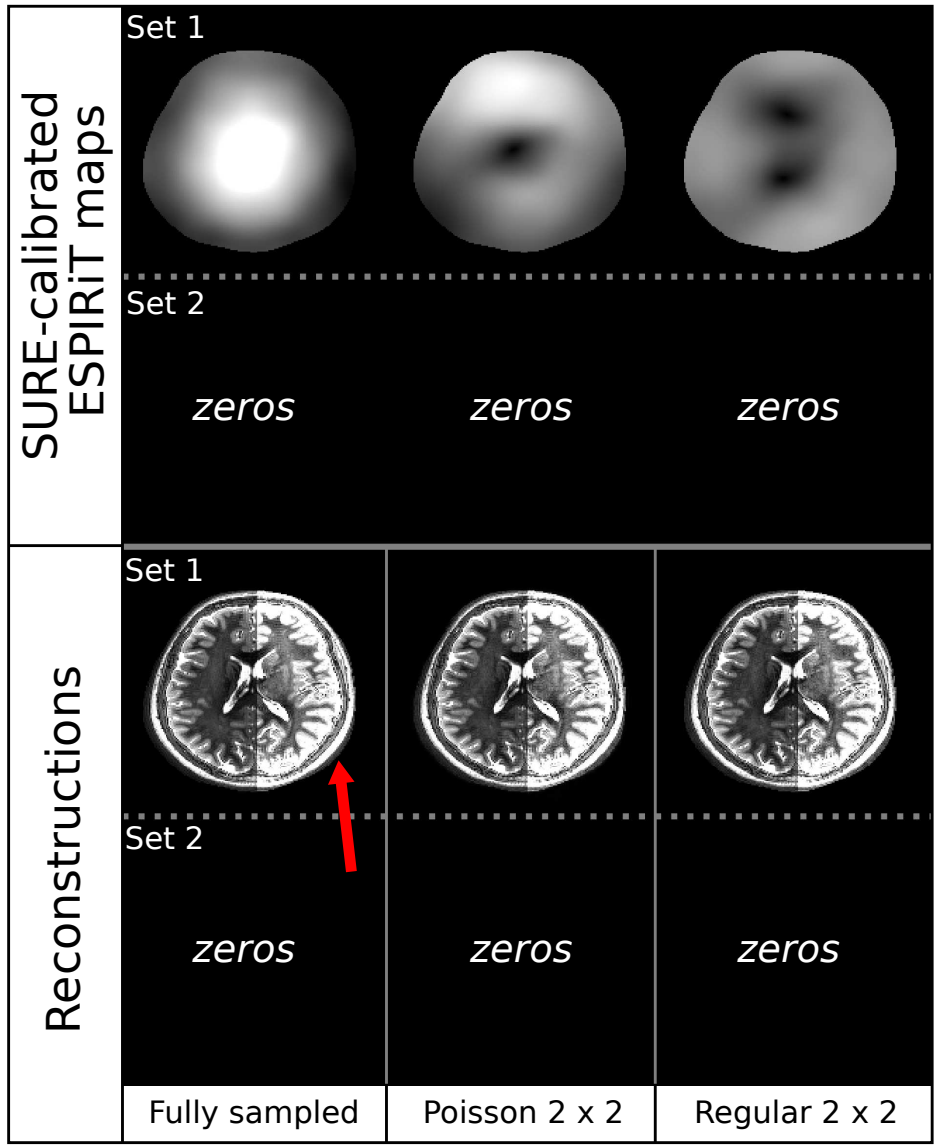}
\caption{
  Experiment (g):
  This figure demonstrates the application of SURE to calibrate ESPIRiT maps on a $T_2$-weighted acquisition.
  The ESPIRiT maps were calibrated using an ACS size of $18 \times 18$ after geometric coil compression.
  These were then used to reconstruct fully sampled k-space, $2\times 2$ retrospectively under-sampled k-space using a
  Poisson mask, and $2\times 2$ retrospectively under-sampled k-space using a regular equi-spaced $2\times 2$ sampling mask.
  Half of the image's contrast has been enhanced to illustrate how well the field of view of the desired object.
  ``Set 1" and ``Set 2" refers the ``soft" SENSE model, where multiple ESPIRiT maps may satisfy the ``$\approx 1$"
  eigenvalue condition. $3$ for the $8$ ESPIRiT maps are shown here for brevity.
}
\label{fig:invivo_t2}
\end{figure}

The results of Experiments (f), (g) and (h) are depicted in Figures \ref{fig:invivo_t1}, \ref{fig:invivo_t2}
and \ref{fig:res_aliasing}, respectively.
Figures \ref{fig:invivo_t1} and \ref{fig:invivo_t2} show how SURE allows ESPIRiT to tightly capture
the field of view of the desired object without attenuating the signal.
Experiment (h) demonstrates that SURE calibration allows ESPIRiT to retain its robustness
to aliasing when the FOV is smaller than object, while still capturing tightly the field of view
of the desired object.

\subsection{Characterization Results}
The result of Experiment (i) is illustrated in Figure \ref{fig:calib_size}.
While a larger calibration size does result in a lower squared error, the advantage is in the
order of $10^5$, which is much smaller than optimizing other parameters depicted in
Figure \ref{fig:espirit_sim}, which are in the order of $10^7$.

The result of Experiment (j) is illustrated in Figure \ref{fig:crop_per_slice}.
While the SURE-based parameters derive ESPIRiT maps that tightly support the FOV of the desired
signal for most slices and is an improvement over the default parameters, the spurious signal outside the human body due to
eye-motion related artifacts in Slice 125 results in ESPIRiT maps that do not non-tightly wrap around the FOV of the desired signal
for both cases.

Experiment (k): The g-factor performance of calibrated ESPIRiT maps as a function of kernel size can be seen in Figure
\ref{fig:g_per_k}.
This suggests that, when using an 8-channel coil, g-factor performance is fairly stable as a function of kernel size.

\section{Discussion}
\label{sec:discussion}

With respect to an exhaustive subspace rank and crop threshold search, SURE as estimated from densely-sampled, high-resolution,
non-accelerated data corresponds well to the true squared error, yielding optimal parameter selection in this full-data case
since SURE is used as a proxy to the true squared error.
The parameters that minimize SURE estimated from ACS correspond well to the optimum parameters that minimize
full-data SURE.
With the assumption that the optimal ESPIRiT maps are derived from parameters that best denoise densely-sampled,
high-resolution, non-accelerated data, the resulting parameter choices are optimal in an expected error sense.

The soft-threshold heuristic is seen to significantly ease computational burden while yielding ESPIRiT maps that perform similarly
to the optimal maps derived from an exhaustive rank search.

Using SURE as a metric to determine ESPIRiT parameters results in consistent performance across different datasets.
In practice, with the parameter ranges mentioned in the in-vivo experiments, the resulting parameter choices
tend to optimize for SNR performance while causing no signal attenuation.

\begin{figure}[ht] \centering
\includegraphics[height=0.75\textheight]{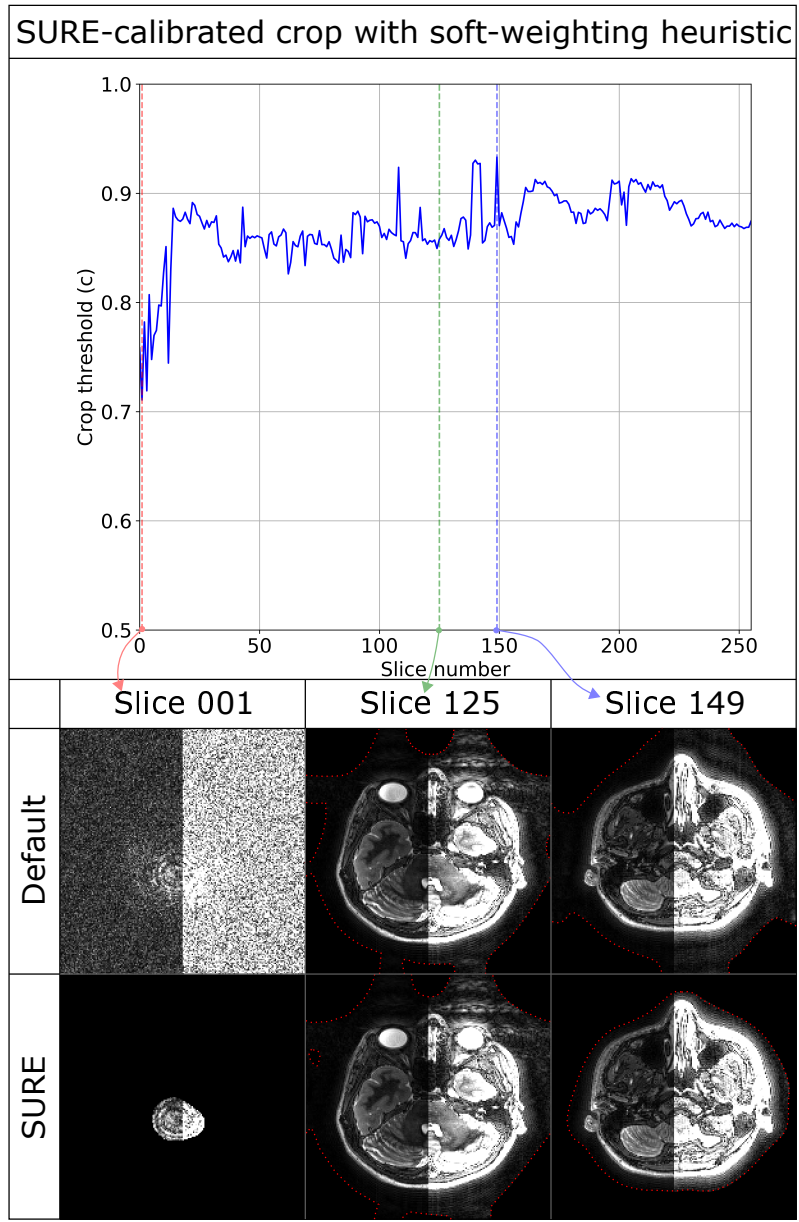}
\caption{
  Experiment (j): SURE-optimal crop threshold per readout slice. If the readout direction is fully-sampled,
  an inverse Fourier transform can be taken and SURE+ESPIRiT can be applied to each readout slice. This will
  allow for ESPIRiT's parameters to vary as a function of noise level per slice. The SURE-calibration used the
  soft-weighting heuristic. The default parameters correspond to a kernel size of $6$ and a WNSVN corresponding
  to the number of singular values greater than $0.001$ times the highest singular value.
  Since SURE tries to avoid data attenuation at all costs, the eye-motion related artifacts in Slice 125 results in
  spurious signal outside the human body. In contrast, the lack of motion in Slice 1 allows SURE to very finely calibrate
  ESPIRiT's support about the FOV of the desired signal.
  The red dotted lines depict the FOV of the calibrated sensitivity maps.
}
\label{fig:crop_per_slice}
\end{figure}

Motion during the scan presents a challenge for estimating accurate auto-calibrated CSM,
and the presented SURE-based parameter optimization can not solve this issue.
In case of motion, the SURE optimized parameters may lead to CSM that extend outside of the images
object as depicted by the eye-motion related artifacts in Slice 125 of Figure \ref{fig:crop_per_slice}.
Since SURE tries to avoid data attenuation at all costs, the ESPIRiT maps calibrated at this slice do not
tightly wrap around the FOV of the desired signal.
This is in contrast to Slice 1 of Figure \ref{fig:crop_per_slice}, where the lack of motion allows SURE and ESPIRiT to
very finely calibrate its support about the FOV of the desired signal.

While the soft-threshold weighting heuristic and line-search based crop threshold search does help
alleviate some of the computational burden, there will always be added cost.
For Experiment (h), ESPIRiT on BART takes $0.416$ seconds using BART's default parameters,
versus $7.112$ seconds using the soft-weighting heuristic and line-search on an Intel(R) Xeon(R) Gold 6138 CPU.
However, this method does offer a robust, data-consistent metric that can adapt to the noise level
for denoising while avoiding signal attenuation.

For practical usage, we recommend a kernel size of $6-8$ with the soft-threshold based weighted 
subspace heuristic and at least $24$ ACS lines.
With respect to the ``soft'' SENSE model described in \cite{ref:uecker}, we recommend using two ESPIRiT maps.
While SURE should help determine the number of ESPIRiT maps as well, it would contribute non-trivially to the
computational burden for minimal gain. If it is known the ACS prescribed FOV is sufficient
for the object, one set of ESPIRiT maps is usually sufficient.

\section{Conclusion}
\label{sec:conclusion}
Using SURE as a metric to determine parameters in ESPIRiT allows for automatic parameter selections.
In-vivo results are consistent with simulation and theoretical results.

\section{Data Availability Statement}
\label{sec:data}

\begin{figure}[ht] \centering
\includegraphics[height=0.7\textheight]{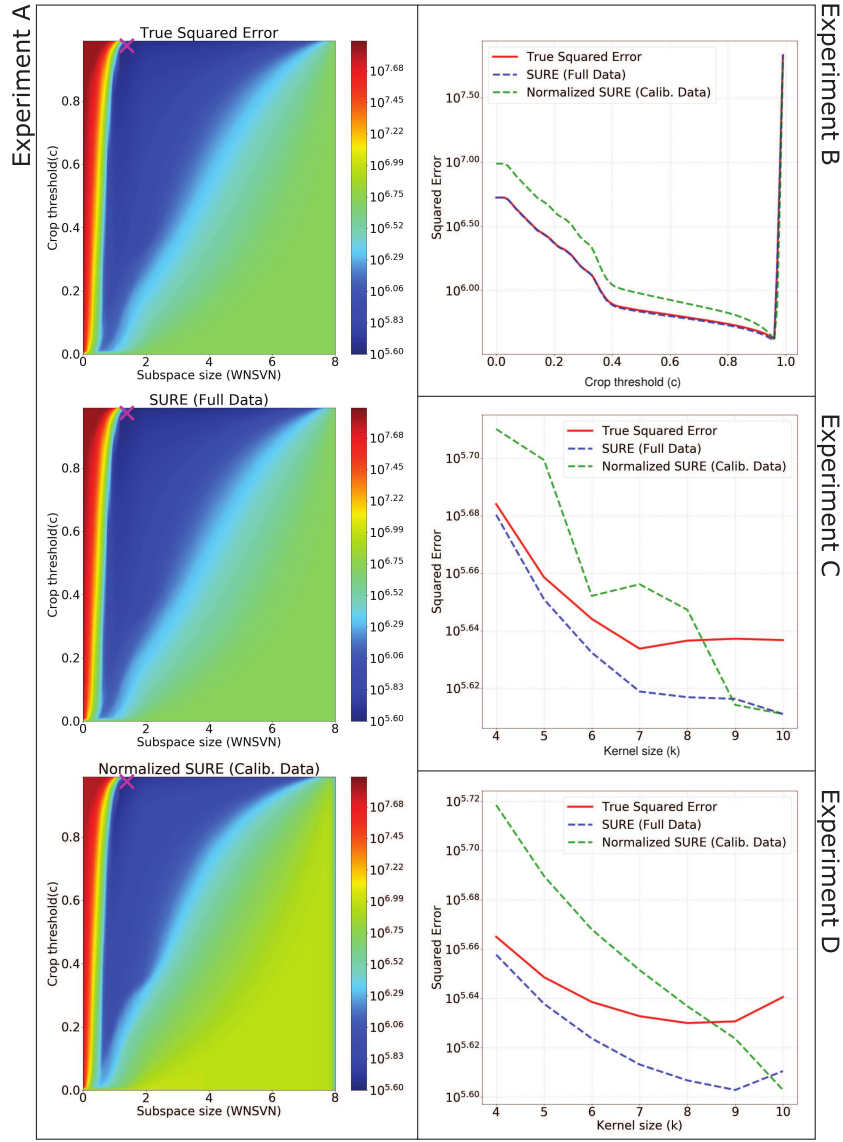}
\caption{
  Experiment (a): with a fixed kernel size of $6$, the true squared error, SURE given densely-sampled,
  high-resolution data and normalized SURE given ACS data is calculated as a function of subspace size
  and eigenvalue crop threshold.The purple crosses show the locations of the respective minimum values.
  Experiment (b): with a fixed kernel size of $6$ along with the soft-threshold based subspace weighting
  heuristic, the true squared error, SURE given densely-sampled, high-resolution data and normalized SURE
  given ACS data is calculated as a function of the eigenvalue crop threshold.
  Experiment (c): True squared error, SURE given densely-sampled, high-resolution data and normalized SURE
  given ACS data is calculated as kernel size, subspace size and eigenvalue crop threshold parameters are varied.
  The minimum given a particular kernel size is taken.
  Experiment (d) is similar to Experiment (c), except that instead of calculating the optimal subspace size given
  the kernel size, the soft-threshold based subspace weighting heuristic is used.
}
\label{fig:espirit_sim}
\end{figure}

\begin{figure}[ht] \centering
\includegraphics[width=0.75\textwidth]{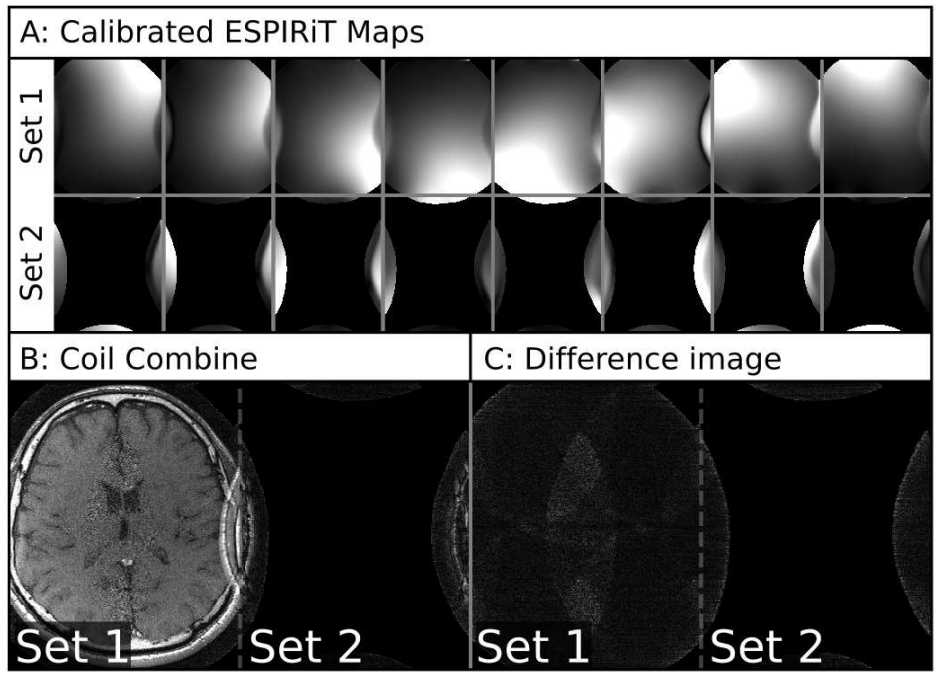}
\caption{
  Experiment (h):
  Calibration With Aliasing Due To FOV Smaller Than Object Result.
  The data is retrospectively under-sampled using an equi-spaced $2\times2$ sampling mask with an ACS size of
  $24\times24$.
  A fixed kernel size of $6$ is used with the soft-threshold based subspace weighting heuristic.
  The eigenvalue crop threshold is varied.
  The data was Fourier transformed to the image domain and noise variance was estimated from a corner of the
  image data that did not contain any desired signal.
  Panel A shows the SURE-calibrated ESPIRiT maps, Panel B shows the $2\times2$ conjugate-gradient reconstruction
  and Panel C shows the difference between the fully sampled image and the $2\times2$ reconstructed image.
  It is seen that the SURE-based parameters selections results in parameters that retain ESPIRiT's robustness
  to image aliasing.
}
\label{fig:res_aliasing}
\end{figure}

\begin{figure}[ht] \centering
\includegraphics[width=0.75\textwidth]{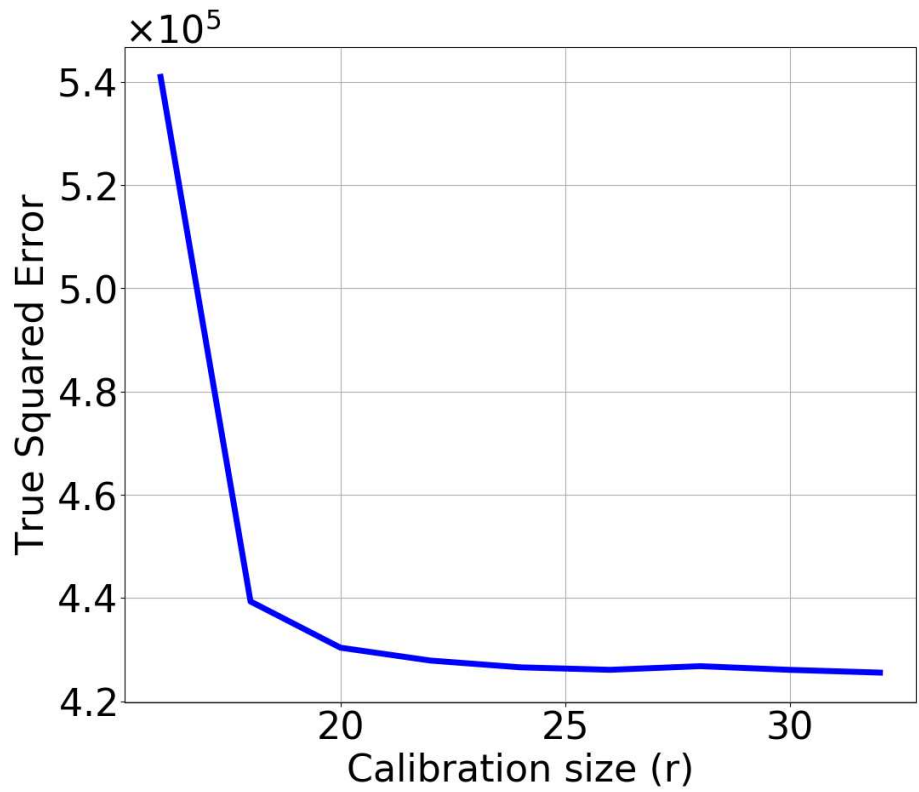}
\caption{
  Experiment (i):
  This figure shows the effect of calibration size on the performance of ESPIRiT maps for a fixed kernel size
  $(k = 8)$ and optimized crop threshold $(c)$ assuming the soft-weighting heuristic. A calibration size of
  $(r)$ implies $(r \times r)$ ACS lines. While a larger calibration size does result in a lower squared error,
  the advantage is on the order of $10^5$, which is much smaller than optimizing other parameters as in
  Experiments (a), (b), (c), and (d), which are on the order of $10^7$.
}
\label{fig:calib_size}
\end{figure}

\begin{figure}[ht] \centering
\includegraphics[width=0.75\textwidth]{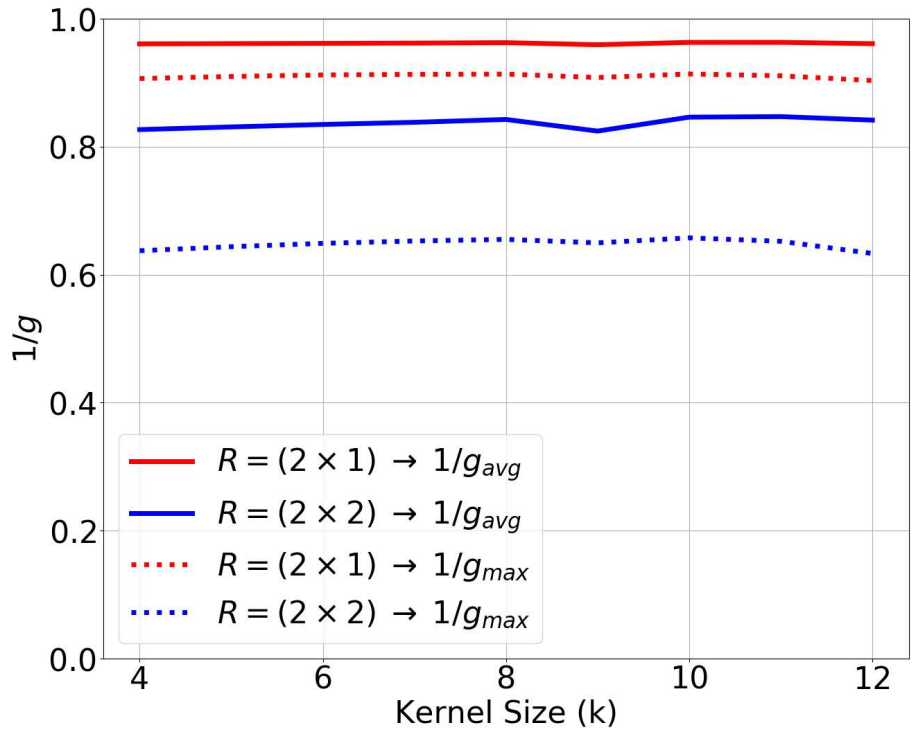}
\caption{
  Experiment (k):
  This figure shows the effect of kernel size on the performance of ESPIRiT maps for assuming SURE-optimized crop 
  threshold $(c)$ and the soft-weighting heuristic.
  The calibration as performed on the same dataset as in Experiments (a), (b), (c) and (d).
  $R = 2 \times 1$ implies equispaced $2 \times 2$ under-sampling in both phase encode directions,
  while $R = 2 \times 1$ implies equispaced $2\times$ under-sampling in one phase encode direction.
  It is seen that, when using an 8-channel coil, g-factor performance appears to be fairly stable as a
  function of kernel size.
}
\label{fig:g_per_k}
\end{figure}

The MATLAB (MathWorks, Natick, MA) code and data that support the findings of the Experiments $(a-e, i, k)$,
used to produce Figures \ref{fig:espirit_sim}, \ref{fig:gfact}, \ref{fig:res_aliasing}, \ref{fig:calib_size},
and \ref{fig:g_per_k}, are openly available in \url{https://github.com/mikgroup/auto-espirit} at DOI:10.5281/zenodo.3679377.
The $C$ code that support the findings of Experiments $(f, g, j)$, used to produce Figures \ref{fig:invivo_t1},
\ref{fig:invivo_t2} and \ref{fig:crop_per_slice}, is openly available as a part of the BART Toolbox in
\url{https://github.com/mrirecon/bart/} at DOI:10.5281/zenodo.3376744, reference \cite{ref:uecker2}.

\bibliography{main}

\begin{thebibliography}{10}

\bibitem{ref:sense}
Pruessmann~KP, Weiger~M, Scheidegger~MB, Boesiger~P et~al.
\newblock {SENSE: Sensitivity Encoding for Fast MRI}.
\newblock {Magnetic Resonance in Medicine} 1999; 42:952--962.

\bibitem{ref:grappa}
Griswold~MA, Jakob~PM, Heidemann~RM, Nittka~M, Jellus~V, Wang~J, Kiefer~B,
  Haase~A.
\newblock {Generalized Autocalibrating Partially Parallel Acquisitions
  (GRAPPA)}.
\newblock {Magnetic Resonance in Medicine} 2002; 47:1202--1210.

\bibitem{ref:uecker}
Uecker~M, Lai~P, Murphy~MJ, Virtue~P, Elad~M, Pauly~JM, Vasanawala~SS,
  Lustig~M.
\newblock {ESPIRiT - An Eigenvalue Approach to Autocalibrating Parallel MRI:
  Where SENSE Meets GRAPPA}.
\newblock Magnetic Resonance in Medicine 2014; 71:990--1001.

\bibitem{ref:stein}
Stein~CM.
\newblock {Estimation of the Mean of a Multivariate Normal Distribution}.
\newblock {The Annals of Statistics} 1981; pp. 1135--1151.

\bibitem{luisier2007new}
Luisier~F, Blu~T, Unser~M.
\newblock {A New SURE Approach to Image Denoising: Interscale Orthonormal
  Wavelet Thresholding}.
\newblock IEEE Transactions on Image Processing 2007; 16:593--606.

\bibitem{zhang1998adaptive}
Zhang~XP, Desai~MD.
\newblock {Adaptive Denoising Based on SURE Risk}.
\newblock IEEE Signal Processing Letters 1998; 5:265--267.

\bibitem{blu2007sure}
Blu~T, Luisier~F.
\newblock {The SURE-LET Approach to Image Denoising}.
\newblock IEEE Transactions on Image Processing 2007; 16:2778--2786.

\bibitem{luisier2008sure}
Luisier~F, Blu~T.
\newblock {SURE-LET Multichannel Image Denoising: Interscale Orthonormal
  Wavelet Thresholding}.
\newblock IEEE Transactions on Image Processing 2008; 17:482--492.

\bibitem{ref:ramani1}
Ramani~S, Blu~T, Unser~M.
\newblock {Monte-Carlo SURE: A Black-box Optimization of Regularization
  Parameters for General Denoising Algorithms}.
\newblock IEEE Transactions on Image Processing 2008; 17:1540--1554.

\bibitem{lustig2010spirit}
Lustig~M, Pauly~JM.
\newblock {SPIRiT: Iterative Self-Consistent Parallel Imaging Reconstruction
  from Arbitrary K-Space}.
\newblock {Magnetic Resonance in Medicine} 2010; 64:457--471.

\bibitem{ref:weller1}
Weller~DS, Ramani~S, Nielsen~JF, Fessler~JA.
\newblock {SURE-based Parameter Selection for Parallel MRI Reconstruction using
  GRAPPA and Sparsity}.
\newblock 2013 IEEE 10th International Symposium on Biomedical Imaging 2013;
  pp. 954--957.

\bibitem{ref:weller2}
Weller~DS, Ramani~S, Nielsen~JF, Fessler~JA.
\newblock {Monte Carlo SURE-based Parameter Selection for Parallel Magnetic
  Resonance Imaging Reconstruction}.
\newblock Magnetic Resonance in Medicine 2014; 71:1760--1770.

\bibitem{xu2014robust}
Xu~L, Feng~Y, Liu~X, Kang~L, Chen~W.
\newblock {Robust GRAPPA Reconstruction using Sparse Multi-kernel Learning with
  Least Squares Support Vector Regression}.
\newblock Magnetic Resonance Imaging 2014; 32:91--101.

\bibitem{majumdar2013calibrationless}
Majumdar~A, Chaudhury~K, Ward~R.
\newblock {Calibrationless Parallel Magnetic Resonance Imaging: A Joint
  Sparsity Model}.
\newblock Sensors 2013; 13:16714--16735.

\bibitem{jin2016general}
Jin~KH, Lee~D, Ye~JC.
\newblock {A General Framework for Compressed Sensing and Parallel MRI using
  Annihilating Filter Based Low-rank Hankel Matrix}.
\newblock IEEE Transactions on Computational Imaging 2016; 2:480--495.

\bibitem{park2012adaptive}
Park~S, Park~J.
\newblock {Adaptive Self-calibrating Iterative GRAPPA Reconstruction}.
\newblock Magnetic Resonance in Medicine 2012; 67:1721--1729.

\bibitem{haldar2016p}
Haldar~JP, Zhuo~J.
\newblock {P-LORAKS: Low-rank Modeling of Local k-space Neighborhoods with
  Parallel Imaging Data}.
\newblock Magnetic Resonance in Medicine 2016; 75:1499--1514.

\bibitem{ref:candes}
Candes~EJ, SingLong~CA, Trzasko~JD.
\newblock {Unbiased Risk Estimates for Singular Value Thresholding and Spectral
  Estimators}.
\newblock IEEE Transactions on Signal Processing 2013; 61:4643--4657.

\bibitem{ying2007joint}
Ying~L, Sheng~J.
\newblock {Joint Image Reconstruction and Sensitivity Estimation in SENSE
  (JSENSE)}.
\newblock Magnetic Resonance in Medicine 2007; 57:1196--1202.

\bibitem{zhao2008iterative}
Zhao~T, Hu~X.
\newblock {Iterative GRAPPA (iGRAPPA) for Improved Parallel Imaging
  Reconstruction}.
\newblock Magnetic Resonance in Medicine 2008; 59:903--907.

\bibitem{chang2012nonlinear}
Chang~Y, Liang~D, Ying~L.
\newblock {Nonlinear GRAPPA: A Kernel Approach to Parallel MRI Reconstruction}.
\newblock Magnetic Resonance in Medicine 2012; 68:730--740.

\bibitem{lyu2018kernl}
Lyu~J, Nakarmi~U, Liang~D, Sheng~J, Ying~L.
\newblock {KerNL: Kernel-based Nonlinear Approach to Parallel MRI
  Reconstruction}.
\newblock IEEE Transactions on Medical Imaging 2018; 38:312--321.

\bibitem{ref:ramani2}
Ramani~S, Liu~Z, Rosen~J, Nielsen~JF, Fessler~JA.
\newblock {Regularization Parameter Selection for Nonlinear Iterative Image
  Restoration and MRI Reconstruction using GCV and SURE-based Methods}.
\newblock IEEE Transactions on Image Processing 2012; 21:3659--3672.

\bibitem{ref:marin}
Marin~A, Chaux~C, Pesquet~JC, Ciuciu~P.
\newblock {Image Reconstruction from Multiple Sensors using Stein's Principle.
  Application to Parallel MRI}.
\newblock 2011 IEEE International Symposium on Biomedical Imaging 2011; pp.
  465--468.

\bibitem{ref:uecker2}
Uecker~M, Ong~F, Tamir~JI, Bahri~D, Virtue~P, Cheng~JY, Zhang~T, Lustig~M.
\newblock {Berkeley Advanced Reconstruction Toolbox}.
\newblock Proc. Intl. Soc. Mag. Reson. Med 2015; 23:2486.

\bibitem{zhang2013coil}
Zhang~T, Pauly~JM, Vasanawala~SS, Lustig~M.
\newblock {Coil Compression for Accelerated Imaging with Cartesian Sampling}.
\newblock Magnetic Resonance in Medicine 2013; 69:571--582.

\bibitem{ref:axler}
Axler~S, ``Linear Algebra Done Right (Undergraduate Texts in Mathematics)''.
  Springer, 2016.

\end{thebibliography}

\section{Appendix}
\label{sec:appendix}

\subsection{Divergence of a Linear Operator (Used in Equations \eqref{eq:surebasic_b} and
\eqref{eq:sureespirit})}
Consider arbitrary $x\in \C^m$, $A \in \C^{m \times m}$ and let $f(x) = Ax$.
\begin{equation}
f(x) = 
\mat{
  A_{11} & \hdots & A_{1m} \\
  \vdots & \ddots & \vdots \\
  A_{m1} & \hdots & A_{mm}
}
\mat{ x_1 \\ \vdots \\ x_m }
\end{equation}
Succinctly, 
\begin{equation}
f_i(x) = \sum_{i=1}^m A_{ij} x_j
\end{equation}
Thus,
\begin{equation} \begin{array}{rl} \left[\text{div}_x f\right](x) 
	&= \sum_{i=1}^m \left[\frac{\partial}{\partial x_i} f_i\right] (x) \\
    &= \sum_{i=1}^m \left[A_{ii}\right] (x) \\
    &= \text{trace}(A)
\end{array}\end{equation}
\subsection{Derivation of Equation \eqref{eq:surebasic}}
Let $x \in \C^m$ be the true, underlying data; $n$ be zero-mean, additive, Gaussian, complex 
noise with the real and imaginary standard deviations of $\frac{\sigma}{\sqrt{2}}$ each; and
$x_{\text{acq}} = x + n$ be acquired data. Let $\pmb{P}$ be the Hermitian symmetric projection operator.

Partition the complex vector space into real and imaginary parts to reduce this problem
to the real case. Let $x_R$ denote the partitioned form of $x$, $\pmb{P}_R$ denote the partitioned
form of $\pmb{P}$ and so on. Then,
\begin{equation}
x_R = \left[\re{x}, \im{x}\right]^T, 
{x_{\text{acq}}}_R = \left[\re{x_{\text{acq}}}, \im{x_{\text{acq}}}\right]^T, 
\pmb{P}_R = \begin{bmatrix}\re{\pmb{P}} & - \im{\pmb{P}}\\\im{\pmb{P}} & \re{\pmb{P}}\end{bmatrix} \text{ etc.}
\end{equation}
This partitioning allows the use \cite{ref:stein} directly. Observe that the noise 
variance in the partitioned case is now $\frac{\sigma^2}{2}$.
\begin{equation}\begin{array}{rl} \text{SURE}_{\pmb{P}}(x_{\text{acq}})
  &= -2m\frac{\sigma^2}{2} + \norm{(\pmb{P} - I)x_{\text{acq}}}_2^2 + 
      2\frac{\sigma^2}{2}\text{trace}(\pmb{P}_R)\\
  &= -m\sigma^2 + \norm{(\pmb{P} - I)x_{\text{acq}}}_2^2 + 2 \sigma^2 \text{trace}(\pmb{P})
\end{array}\end{equation}
Using the fact that the eigenvalues of $\pmb{P}$ are real (since $\pmb{P}$ is Hermitian symmetric) and that the
trace of a matrix is independent from basis of representation \cite{ref:axler}, it follows:
\begin{equation}
\text{trace}(\pmb{P}_R) = 2 \text{trace}(\pmb{P})
\end{equation}

\clearpage

\begin{figure}[ht] \centering
\includegraphics[width=0.95\textwidth]{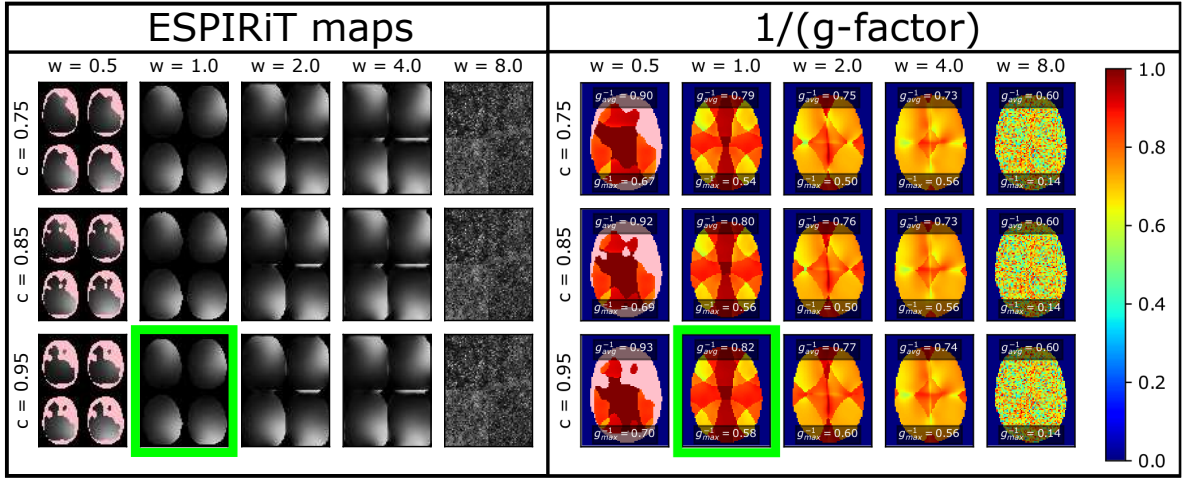}
\caption{
    Variability in ESPIRiT maps is exemplified by varying the signal subspace size $(w)$ in WNSVN and eigenvalue
    crop threshold $(c)$.
    These parameters are explained in the theory section. 
    The mean g-factor $g_{\text{avg}}$ and maximum g-factor $g_{\text{max}}$ within the field of view of the
    desired object is used to quantify the performance of the ESPIRiT maps.
    $1/g_{\text{avg}}$ and $1/g_{\text{max}}$ are listed at top and bottom of the g-factor maps respectively.
    The g-factor is calculated assuming equispaced $2\times 2$ sub-sampling in the phase encode directions.
    ESPIRiT maps that do not attenuate signal but have a high $1/g_{\text{avg}}$ and $1/g_{\text{max}}$ are
    desirable.
    Light pink mark regions of signal attenuation.
    In this figure, the best example is marked with a green box.
}
\label{fig:variability}
\end{figure}
\clearpage

\clearpage

\clearpage

\begin{figure}[ht] \centering
\includegraphics[width=0.75\textwidth]{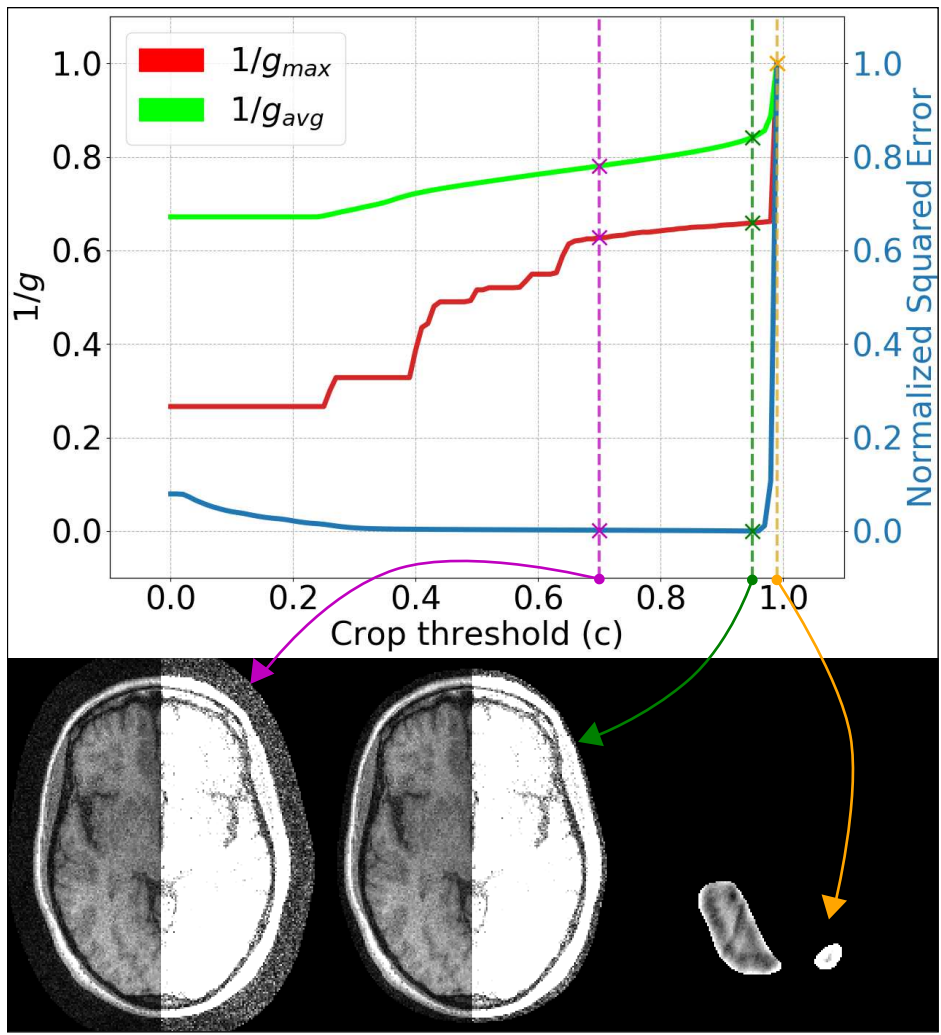}
\caption{
  This figure shows how g-factor performance varies as a function of the Crop Threshold ($c$) when using the
  soft-weighting heuristic. The g-factor was calculated
  for voxels within the field of view of the brain
  using the same $2 \times 2$ equi-spaced sampling pattern
  described in Figure \ref{fig:variability}. Note that for all values of $(c)$ to the left of the green line
  (marked $B$), an increase in $(c)$ results in a decrease in squared error as well as better g-factor 
  performance. However, for values of $(c)$ to the right of $B$, the resulting maps attenuate the desired
  signal. This is reflected by the increase in squared error. The purple line $A$ reflects a safe value of
  the crop threshold, $B$ reflects the optimal value in terms of minimum squared error and $C$ represents the
  best g-factor but at the cost of a lot of signal attenuation. The optimality of $B$ is also reflected in
  how well the field of view of the desired signal is captured. Half of the image's contrast has been enhanced
  to illustrate this point.
}
\label{fig:gfact}
\end{figure}
\clearpage

\clearpage

\clearpage

\clearpage

\clearpage

\clearpage

\clearpage

\textbf{\large List of Figures}
\begin{enumerate}
\item
    Variability in ESPIRiT maps is exemplified by varying the signal subspace size $(w)$ in WNSVN and eigenvalue
    crop threshold $(c)$.
    These parameters are explained in the theory section. 
    The mean g-factor $g_{\text{avg}}$ and maximum g-factor $g_{\text{max}}$ within the field of view of the
    desired object is used to quantify the performance of the ESPIRiT maps.
    $1/g_{\text{avg}}$ and $1/g_{\text{max}}$ are listed at top and bottom of the g-factor maps respectively.
    The g-factor is calculated assuming equispaced $2\times 2$ sub-sampling in the phase encode directions.
    ESPIRiT maps that do not attenuate signal but have a high $1/g_{\text{avg}}$ and $1/g_{\text{max}}$ are
    desirable.
    Light pink mark regions of signal attenuation.
    In this figure, the best example is marked with a green box.
\item
    This summarizes the technique of generating ESPIRiT maps that are used as 
    SENSE-like projection operators. (a) A kernel is swept through the calibration region to
    construct the calibration matrix (which consequently has block Hankel structure). (b) The
    SVD of the calibration matrix is taken and (c) the right singular vectors corresponding to the
    largest singular values are reshaped into k-space kernels. (d) The inverse Fourier transform
    of these kernels are taken, followed by (e) the eigenvalue decomposition of each pixel along the
    coil dimension in the image domain. The eigenvectors corresponding to eigenvalues 
    ``$\approx 1$'' are used to construct the ESPIRiT operator. (f) depicts how ESPIRiT maps
    can be used as a projection operator to denoise data in the ideal case. (g) Consequently, 
    the null space of the ideal projection operator would contain only noise.
\item
    Experiment (a): with a fixed kernel size of $6$, the true squared error, SURE given densely-sampled,
    high-resolution data and normalized SURE given ACS data is calculated as a function of subspace size
    and eigenvalue crop threshold. The purple crosses show the locations of the respective minimum values.
    Experiment (b): with a fixed kernel size of $6$ along with the soft-threshold based subspace weighting
    heuristic, the true squared error, SURE given densely-sampled, high-resolution data and normalized SURE
    given ACS data is calculated as a function of the eigenvalue crop threshold.
    Experiment (c): True squared error, SURE given densely-sampled, high-resolution data and normalized SURE
    given ACS data is calculated as kernel size, subspace size and eigenvalue crop threshold parameters are varied.
    The minimum given a particular kernel size is taken.
    Experiment (d) is similar to Experiment (c), except that instead of calculating the optimal subspace size given
    the kernel size, the soft-threshold based subspace weighting heuristic is used.
\item
    This figure shows how g-factor performance varies as a function of the Crop Threshold ($c$) when using the
    soft-weighting heuristic. The g-factor was calculated
    for voxels within the field of view of the brain
    using the same $2 \times 2$ equi-spaced sampling pattern
    described in Figure \ref{fig:variability}. Note that for all values of $(c)$ to the left of the green line
    (marked $B$), an increase in $(c)$ results in a decrease in squared error as well as better g-factor 
    performance. However, for values of $(c)$ to the right of $B$, the resulting maps attenuate the desired
    signal. This is reflected by the increase in squared error. The purple line $A$ reflects a safe value of
    the crop threshold, $B$ reflects the optimal value in terms of minimum squared error and $C$ represents the
    best g-factor but at the cost of a lot of signal attenuation. The optimality of $B$ is also reflected in
    how well the field of view of the desired signal is captured. Half of the image's contrast has been enhanced
    to illustrate this point.
\item
    Experiment (f):
    This figure demonstrates the application of SURE to calibrate ESPIRiT maps on a $T_1$-weighted acquisition
    where roughly $50\%$ of the total phase encode points are acquired, which included an ACS size of $18\times 18$.
    Half of the image's contrast has been enhanced to illustrate how well the field of view of the desired object
    has been captured.
    ``Set 1'' and ``Set 2'' refers the ``soft'' SENSE model, where multiple ESPIRiT maps may satisfy the ``$\approx 1$''
    eigenvalue condition.
    $2$ of the $8$ calibrated ESPIRiT maps are depicted in this figure for brevity.
\item
    Experiment (g):
    This figure demonstrates the application of SURE to calibrate ESPIRiT maps on a $T_2$-weighted acquisition.
    The ESPIRiT maps were calibrated using an ACS size of $18 \times 18$ after geometric coil compression.
    These were then used to reconstruct fully sampled k-space, $2\times 2$ retrospectively under-sampled k-space using a
    Poisson mask, and $2\times 2$ retrospectively under-sampled k-space using a regular equispaced $2\times 2$ sampling mask.
    Half of the image's contrast has been enhanced to illustrate how well the field of view of the desired object.
    ``Set 1'' and ``Set 2'' refers the ``soft'' SENSE model, where multiple ESPIRiT maps may satisfy the ``$\approx 1$''
    eigenvalue condition. $3$ for the $8$ ESPIRiT maps are shown here for brevity.
\item
    Experiment (h):
    Calibration With Aliasing Due To FOV Smaller Than Object Result.
    The data is retrospectively under-sampled using an equi-spaced $2\times2$ sampling mask with an ACS size of
    $24\times24$.
    A fixed kernel size of $6$ is used with the soft-threshold based subspace weighting heuristic.
    The eigenvalue crop threshold is varied.
    The data was Fourier transformed to the image domain and noise variance was estimated from a corner of the
    image data that did not contain any desired signal.
    Panel A shows the SURE-calibrated ESPIRiT maps, Panel B shows the $2\times2$ conjugate-gradient reconstruction
    and Panel C shows the difference between the fully sampled image and the $2\times2$ reconstructed image.
    It is seen that the SURE-based parameters selections results in parameters that retain ESPIRiT's robustness
    to image aliasing.
\item
    Experiment (i):
    This figure shows the effect of calibration size on the performance of ESPIRiT maps for a fixed kernel size
    $(k = 8)$ and optimized crop threshold $(c)$ assuming the soft-weighting heuristic. A calibration size of
    $(r)$ implies $(r \times r)$ ACS lines. While a larger calibration size does result in a lower squared error,
    the advantage is on the order of $10^5$, which is much smaller than optimizing other parameters as in
    Experiments (a), (b), (c), and (d), which are on the order of $10^7$.
\item
    Experiment (j): SURE-optimal crop threshold per readout slice. If the readout direction is fully-sampled,
    an inverse Fourier transform can be taken and SURE+ESPIRiT can be applied to each readout slice. This will
    allow for ESPIRiT's parameters to vary as a function of noise level per slice. The SURE-calibration used the
    soft-weighting heuristic. The default parameters correspond to a kernel size of $6$ and a WNSVN corresponding
    to the number of singular values greater than $0.001$ times the highest singular value.
    Since SURE tries to avoid data attenuation at all costs, the eye-motion related artifacts in Slice 125 results in
    spurious signal outside the human body. In contrast, the lack of motion in Slice 1 allows SURE to very finely calibrate
    ESPIRiT's support about the FOV of the desired signal.
    The red dotted lines depict the FOV of the calibrated sensitivity maps.
\item
    Experiment (k):
    This figure shows the effect of kernel size on the performance of ESPIRiT maps for assuming SURE-optimized crop 
    threshold $(c)$ and the soft-weighting heuristic.
    The calibration as performed on the same dataset as in Experiments (a), (b), (c) and (d).
    $R = 2 \times 1$ implies equispaced $2 \times 2$ under-sampling in both phase encode directions,
    while $R = 2 \times 1$ implies equispaced $2\times$ under-sampling in one phase encode direction.
    It is seen that, when using an 8-channel coil, g-factor performance appears to be fairly stable as a
    function of kernel size.
\end{enumerate}

\end{document}